\documentclass[12pt,preprint,showpacs,preprintnumbers]{revtex4}
\usepackage{amsfonts}
\usepackage{float}
\usepackage{amsmath}
\usepackage{amssymb}
\usepackage{hyperref}
\usepackage{color}
\usepackage{graphicx}
\usepackage{amssymb}
\usepackage{amsmath}
\usepackage{graphicx}
\usepackage{dcolumn}
\usepackage{bm}
\usepackage{epsfig}
\usepackage[T1]{fontenc}
\usepackage{ae,aecompl}

\begin{document}
\title{Thermo-hydrodynamic non-equilibrium effects on compressible Rayleigh-Taylor instability}
\author{ Huilin Lai$^{1,2}$, Aiguo Xu$^{1,3}$\footnote{Corresponding author: Xu\_Aiguo@iapcm.ac.cn}, Guangcai Zhang$^{1}$\footnote{Corresponding author: Zhang\_Guangcai@iapcm.ac.cn}, Yanbiao Gan$^{1,4}$,
Yangjun Ying$^1$, Sauro Succi$^5$}
\affiliation{$^1$ National Key Laboratory of Computational Physics, Institute of Applied Physics and Computational Mathematics, P. O. Box 8009-26, Beijing 100088, China \\
$^2$ School of Mathematics and Computer Science, Fujian Normal University, Fuzhou 350007, China\\
$^3$ Center for Applied Physics and Technology, MOE Key Center for High
Energy Density Physics Simulations, College of Engineering, Peking
University, Beijing 100871, China \\
$^4$ North China Institute of Aerospace Engineering, Langfang 065000, China \\
$^5$ Istituto Applicazioni Calcolo, CNR, Viale del Policlinico 137, 00161 Roma, Italy.
}
\date{\today }

\begin{abstract}
The effects of compressibility on Rayleigh-Taylor instability (RTI) are investigated by inspecting the interplay between thermodynamic and hydrodynamic non-equilibrium phenomena (TNE, HNE, respectively) via a discrete Boltzmann model (DBM).
Two effective approaches are presented, one tracking the evolution of the \emph{local} TNE effects and the other focussing on the evolution of the mean temperature of the fluid,
to track the complex interfaces separating the bubble and the spike regions of the flow.
It is found that, both the compressibility effects and the \emph{global} TNE intensity show opposite trends in the initial and the later stages of the RTI. Compressibility delays the initial stage of RTI and accelerates the later stage.
Meanwhile, the TNE characteristics are generally enhanced by the compressibility, especially in the later stage.
The global or mean thermodynamic non-equilibrium indicators provide physical criteria to discriminate between the two stages of the RTI.
\end{abstract}

\maketitle

\section{Introduction}

Rayleigh-Taylor instability (RTI) occurs at the interface between two fluids with different
densities, subjected to an acceleration directed from the lower density fluid to the higher density one.
A typical case is a heavy fluid resting on the top of a lighter one in the presence of a gravitational field. Under such conditions, density perturbations at the interface grow in time under the effect of gravity.
The first detailed study of this instability was
conducted by Rayleigh \cite{Rayleigh} in the 1880s. Later the first study was extended to accelerated fluids by Taylor \cite{Taylor} in 1950. The first experiment was performed by Lewis in the evolution of an unstable air-water interface \cite{Lewis}. Another experiment by Emmons \emph{et al}. confirmed these findings \cite{Emmons}.
Such an instability plays a prominent role in many natural and industrial processes, such as devices for sustainable energy production, say turbines \cite{JuYG} and inertial-confinement fusion (ICF) \cite{ICF}, type-la supernovae \cite{supernovae}, hot-wire diagnostics \cite{Kraft}, quantum magnetized plasmas \cite{WLF2014}, colloidal mixtures \cite{soft2011}, etc.

In the above-mentioned fields, the compressibility effects on RTI are essential, and even dominate \cite{Bernstein-Book,Livescu2,Benzi01,Benzi02}, deserving careful investigation.
In fact, many theoretical and numerical studies have been performed, especially on the initial linear stage \cite{Jin,Hoshoudy,Livescu3,Lafay,He-Hu,Xue-Ye,Gauthier}. In those studies, the compressibility effects on RTI growth rate are generally probed via changing the ratio of specific heats and the equilibrium pressure at the interface. Specifically, in 2007, Lafay \emph{et al.} found that, in isothermal case, the stratification has a stabilizing effect while the compressibility has a destabilizing effect for two miscible viscous and compressible fluids \cite{Lafay}. In 2008, He \emph{et al.} reported that, in an inviscid case, the influences of the ratio of specific heats are as below. It mitigates the RTI when the upper heavy fluid is more compressible, while it enhances the RTI when the lower fluid is more compressible \cite{He-Hu}. In 2010, Ye \emph{et al.} demonstrated that the compressibility has destabilizing effects for inviscid compressible fluid with exponentially variable density profile \cite{Xue-Ye}.

Although the compressibility effects have been studied extensively,
 several fundamental problems remain open, such as the nonequilibrium effects in RTI, especially for the case of increasing compressibility \cite{Gauthier,Abarzhi,Livescu}.
For the case with strong compressibility, the interfacial dynamics becomes more complicated as the RTI unfolds, resulting in very substantial gradient forces ($\nabla \rho $,$\nabla \mathbf{u}$ and $\nabla T$) around the interfaces and very pronounced thermodynamic non-equilibrium (TNE) effects, where
$\rho $,$\mathbf{u}$ and $T$ are the local density, flow velocity and temperature, respectively.
The more pronounced the compressibility, the more complex the interfaces and the TNE effects as well. It is known that the Navier-Stokes model falls short of  describing the complex interfaces and TNE effects \cite{Succi-Book,Review2012,XuYan2013,XuGan2013,XuLin2014,XuLin2016,XuLin2015PRE,XuGan2015}. At the same time, molecular dynamics and Monte Carlo simulations can not access macroscopic spatial-temporal scales of interest
at affordable computational cost. Under such conditions, a kinetic  approach  based on a suitably simplified model Boltzmann equation is preferable.

As a special discretization of the Boltzmann equation, the lattice Boltzmann (LB) method has achieved great success in various complex flows \cite{Succi-Book,Review2012,Ottaviani,Succi2015,LiQ2,LiQ3,WangY,Yeomans-group1,Yeomans-group2,Yeomans-group3,Zhangyh-group1,Zhangyh-group2,Zhangyh-group3}. The LB applications in RTI can be classified into two groups, RTI in incompressible flows \cite{Clavin,Gunstensen,Nie,He1,He2,ZhangRY,Clark,LiQ1,LiuGuo,LiangShi,Livescu,Abarzhi} and in compressible flows \cite{Sbragaglia1,Sbragaglia2,Sbragaglia3}. In these studies, the LB method appears as an effective numerical scheme to solve the traditional hydrodynamic models.
In recent works \cite{Review2012,XuYan2013,XuGan2013,XuLin2014,XuLin2016,XuLin2015PRE,XuGan2015}, the LB method was developed to probe  the trans- and supercritical fluid behaviors or both the hydrodynamic non-equilibrium (HNE) and TNE behaviours, which has brought some new physical insights into the fundamental mechanisms of the system. Physically, such an extended LB kinetic model or discrete Boltzmann model (DBM) is roughly equivalent to a hydrodynamic model supplemented by a coarse-grained model of the TNE behaviours \footnote{By coarse-grained, we imply that since only a finite number of kinetic models are retained, the fine-structure of non-equilibrium phenomena cannot be fully resolved by the simulation.}.
The DBM has been applied in the combustion system, phase separation system, and compressible flow system with shocks \cite{Review2012,XuYan2013,XuGan2013,XuLin2014,XuLin2016,XuLin2015PRE,XuGan2015}, but still not
in the RTI system.
In this work, we further extend the DBM to investigate both the HNE and TNE behaviours in the compressible RTI system. Compared with previous studies on RTI, besides the compressibility effects, the interplays of various  non-equilibrium behaviours are our main concerns.

The rest of the paper is structured as follows. In Sec. II, we first briefly review the DBM used in this work, then show the basic evolutions of so called ``single" mode RTI and its TNE effects. A new front-tracking scheme based on the TNE property is presented in the same section.  The effects of compressibility on RTI are studied in detail in Sec. III. Sec. IV summarizes and concludes the present paper.

\section{Evolutions of RTI and its TNE characterizations}

\subsection{Discrete Boltzmann Model}

Instead of using the traditional Navier-Stokes equations, in this work the compressible RTI system is described by the following discrete Boltzmann equation with Bhatnagar-Gross-Krook model \cite{Nie,He1},
\begin{equation}\label{eq1}
\dfrac{\partial f_i}{\partial t}+\mathbf{v}_{i}\cdot\dfrac{\partial f_i}{\partial \mathbf{r}}-\dfrac{\mathbf{a}\cdot(\mathbf{v}_{i}-\mathbf{u})}{T}f_{i}^{eq}=-\frac{1}{\tau}( f_i-f_i^{eq}),
\end{equation}
where $f_i(\mathbf{r},\mathbf{v}_{i},t)$ is the discrete distribution function, $\mathbf{r}$ the spatial coordinate, $t$ the time, $\mathbf{v}_{i}$ the discrete velocity model and $i=1,2,\cdots,N$ the direction of the discrete velocity model. $\mathbf{u}$ is the macroscopic velocity, $\mathbf{a}$ an external body force, $\tau$ the relaxation time, and $f_i^{eq}$ is the equilibrium distribution function.

Following the ideas presented in Refs. \cite{Review2012,XuYan2013,XuGan2013,XuLin2014,XuLin2016,XuLin2015PRE,XuGan2015}, we use
\begin{eqnarray}\label{eq2}
\boldsymbol{\Delta}^*_{m,n} &=&\mathbf{M}^*_{m,n}(f_i)-\mathbf{M}^*_{m,n}(f_i^{eq}),
\end{eqnarray}
to describe the TNE effects, where $\mathbf{M}^*_{m,n}$ represent the kinetic central moments, in which the variable $\mathbf{v}$ in Eqs. (\ref{mo4})-(\ref{mo7}) (see the Appendix) is replaced by $\mathbf{v}^*=\mathbf{v}-\mathbf{u}$. ${\mathbf M}^*_{2}$, ${\mathbf M}^*_{3}$, ${\mathbf M}^*_{3,1}$ and ${\mathbf M}^*_{4,2}$ are associated with the Non-Organized Momentum Flux (NOMF), Non-Organized Stress Flux (NOSF), Non-Organized Energy Flux (NOEF), and Flux of NOEF, respectively. Here, high order kinetic moments reflect the molecular individualism on top of organised collective motion, which we conventionally label as non-organised (NO) modes.

In our simulations, we study the spatiotemporal evolutions of a single-component fluid when initially prepared on the hydrostatic unstable equilibrium, i.e., with a cold uniform region in the top half and a hot uniform region in the bottom half. In the two half volumes, we fix two different homogeneous temperatures, with the corresponding hydrostatic density profiles \cite{Sbragaglia2}. We consider the lower and upper borders are kept far from the perturbation interface during the process of RTI, so there is no heat change in the lower and upper borders. Thus, adiabatic and non-slip boundary conditions are applied at the top and bottom walls and periodic boundary conditions on the horizontal boundaries. Specifically, at the top and bottom boundaries, we set the velocity to be zero. The forward Euler finite difference scheme and the nonoscillatory nonfree dissipative scheme \cite{NND} are used to discretize the temporal and spatial derivatives, respectively.

\subsection{Evolutions of RTI}

The starting configuration of RTI is a compressible flow in a 2D domain $[-d/2,d/2]\times [-2d,2d]$. We consider two layers of the fluids at rest in the constant gravity field with the initial position of the interface $y_c(x)=y_0 \cos(k x)$, where $y_0=0.05d$, the wave number $k=2\pi/\lambda$, and the $\lambda=d$ is the wavelength of the perturbation. The temperatures of two half volumes are initially constants and each of the half part is in hydrostatic equilibrium
\begin{equation}\label{eq3}
\partial_y P_0(y)=-g\rho_0(y).
\end{equation}
So the initial hydrostatic unstable configuration is given by
\begin{equation}\label{eq4}
\left\{
\begin{array}{l}
T_0(y)=T_u,\rho_0(y)=\dfrac{P_0}{T_u}\exp{\big[\dfrac{g}{T_u}\big(2d-y\big)\big]},y>y_c(x), \\[8pt]
T_0(y)=T_d,\rho_0(y)=\dfrac{P_0}{T_d}\exp\big[\dfrac{g}{T_u}\big(2d-y_c(x)\big)
\\[8pt]
-\dfrac{g}{T_d}\big(y-y_c(x)\big)\big],y<y_c(x),
\end{array}
\right.
\end{equation}
where $P_0$ is the initial pressure at the top boundary. $T_u$ and $T_d$ represent the initial temperature of the upper half part and the lower half part, respectively. Under this condition, we have the same pressure at the interface
\begin{equation}\label{eq5}
\rho_u T_u=\rho_d T_d,
\end{equation}
where $\rho_u$ and $\rho_d$ are the densities of the upper and the lower grid aside the interface. Then the initial Atwood number can be defined as \cite{Sbragaglia2}
\begin{equation}\label{eq6}
At=\dfrac{\rho_u-\rho_d}{\rho_u+\rho_d}=\dfrac{T_d-T_u}{T_d+T_u}.
\end{equation}

\begin{figure*}[!ht]
\center {
{\epsfig{file=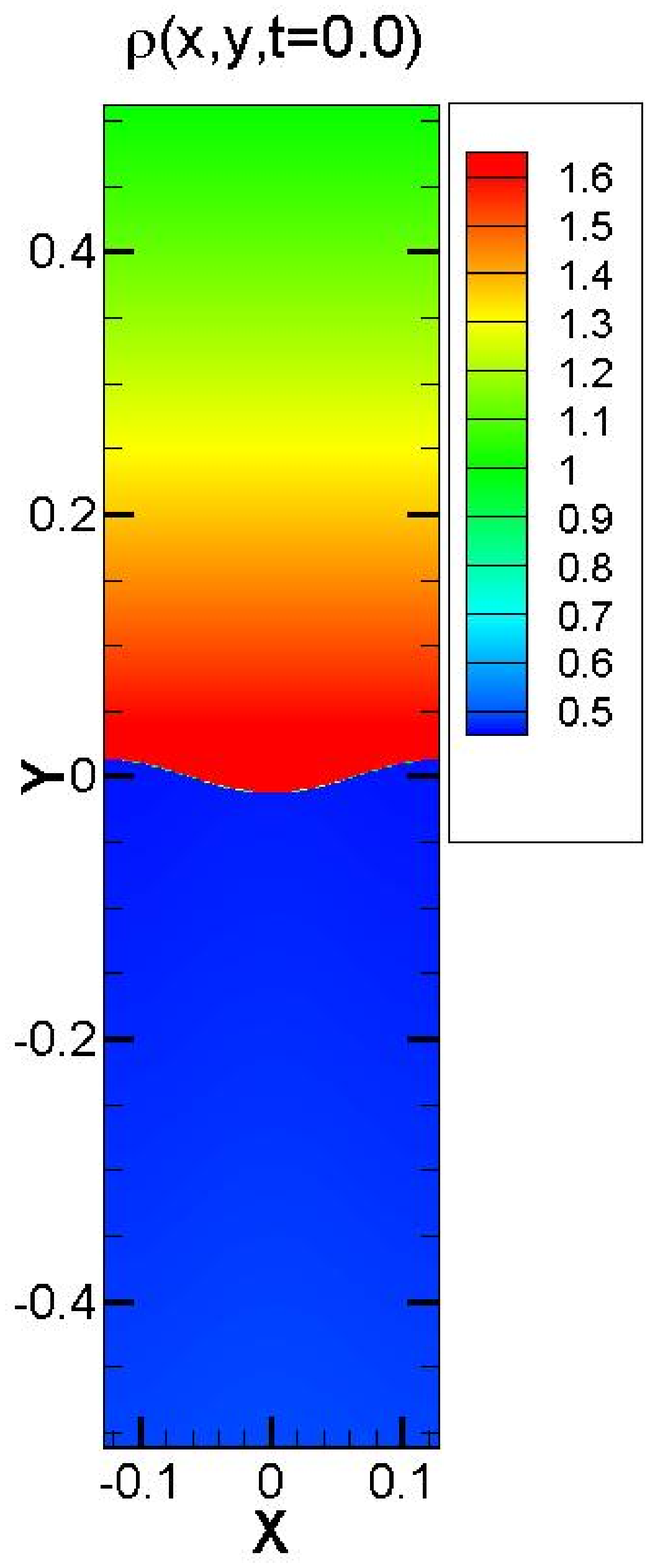,bbllx=22pt,bblly=22pt,bburx=320pt,bbury=742pt,
width=0.165\textwidth,clip=}}\hspace{0.5cm}
{\epsfig{file=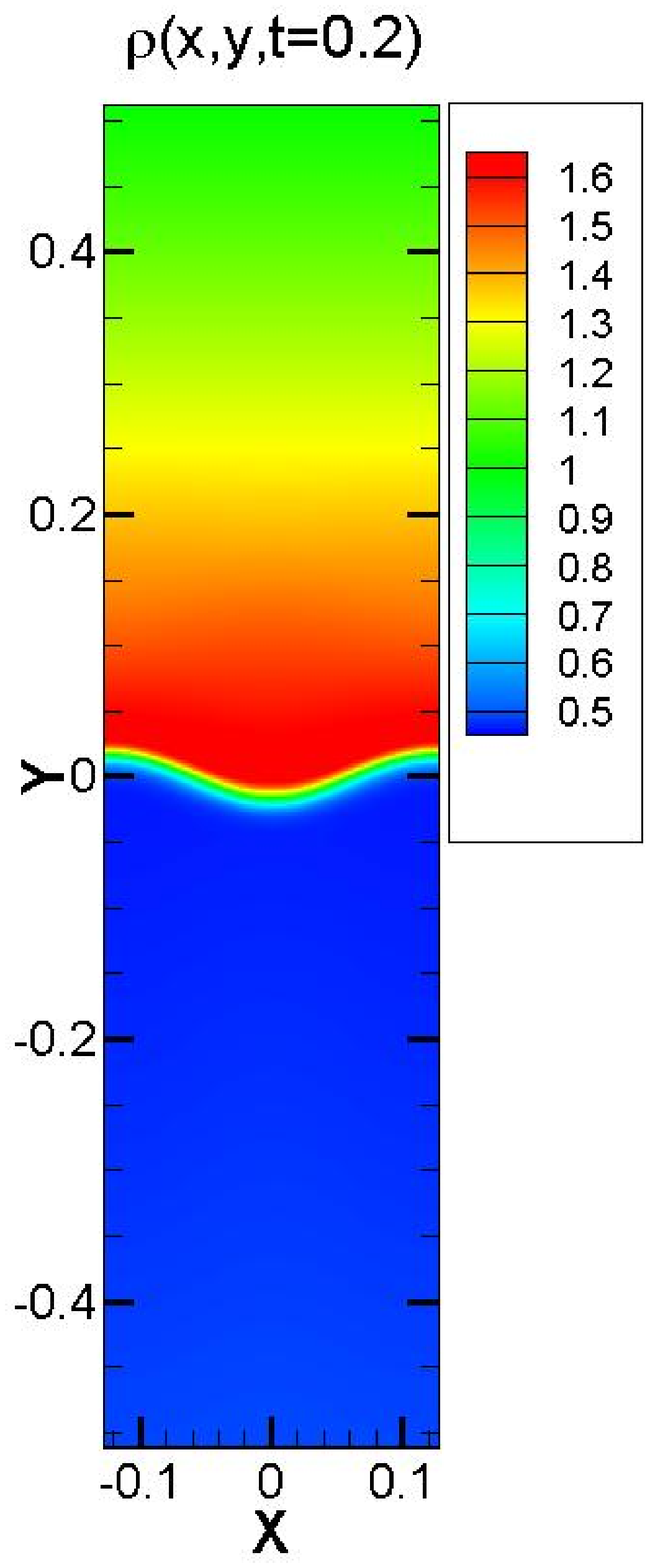,bbllx=22pt,bblly=22pt,bburx=320pt,bbury=742pt,
width=0.165\textwidth,clip=}}\hspace{0.5cm}
{\epsfig{file=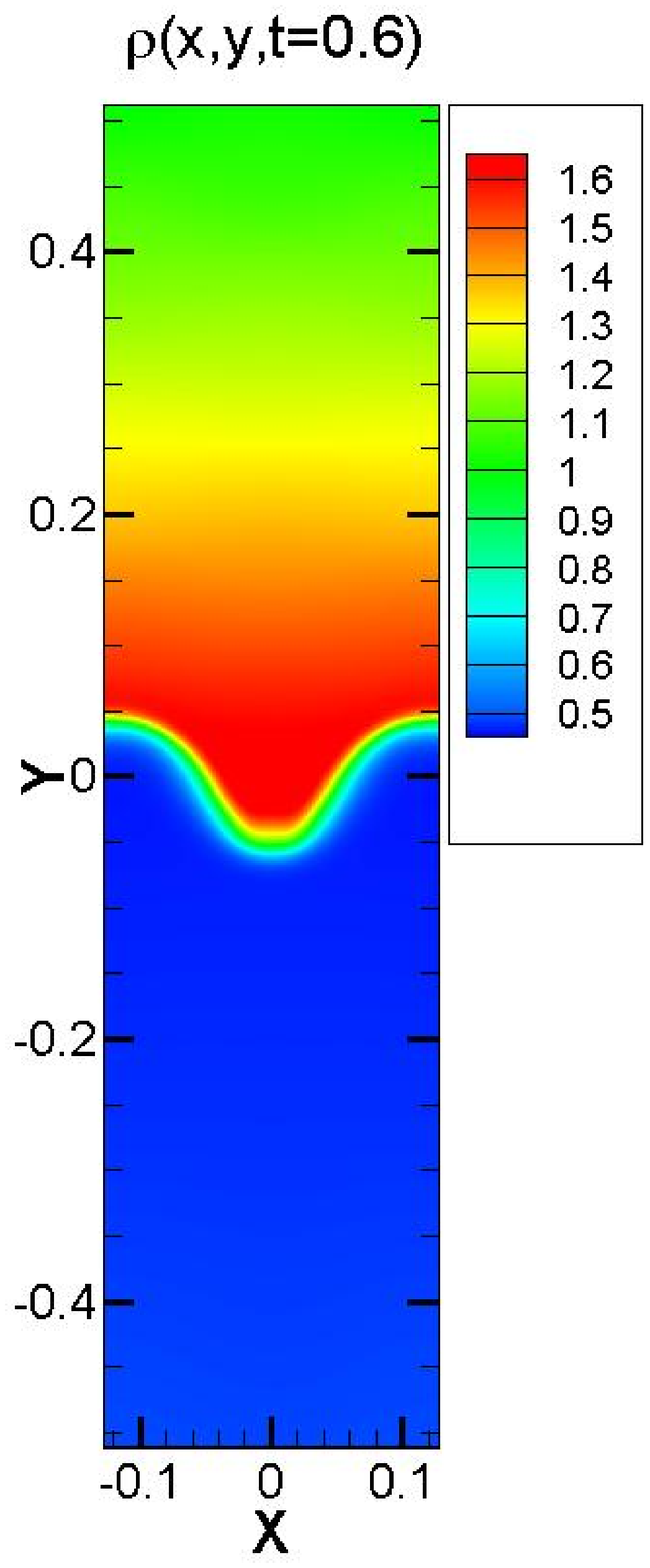,bbllx=22pt,bblly=22pt,bburx=320pt,bbury=742pt,
width=0.165\textwidth,clip=}}\hspace{0.5cm}
{\epsfig{file=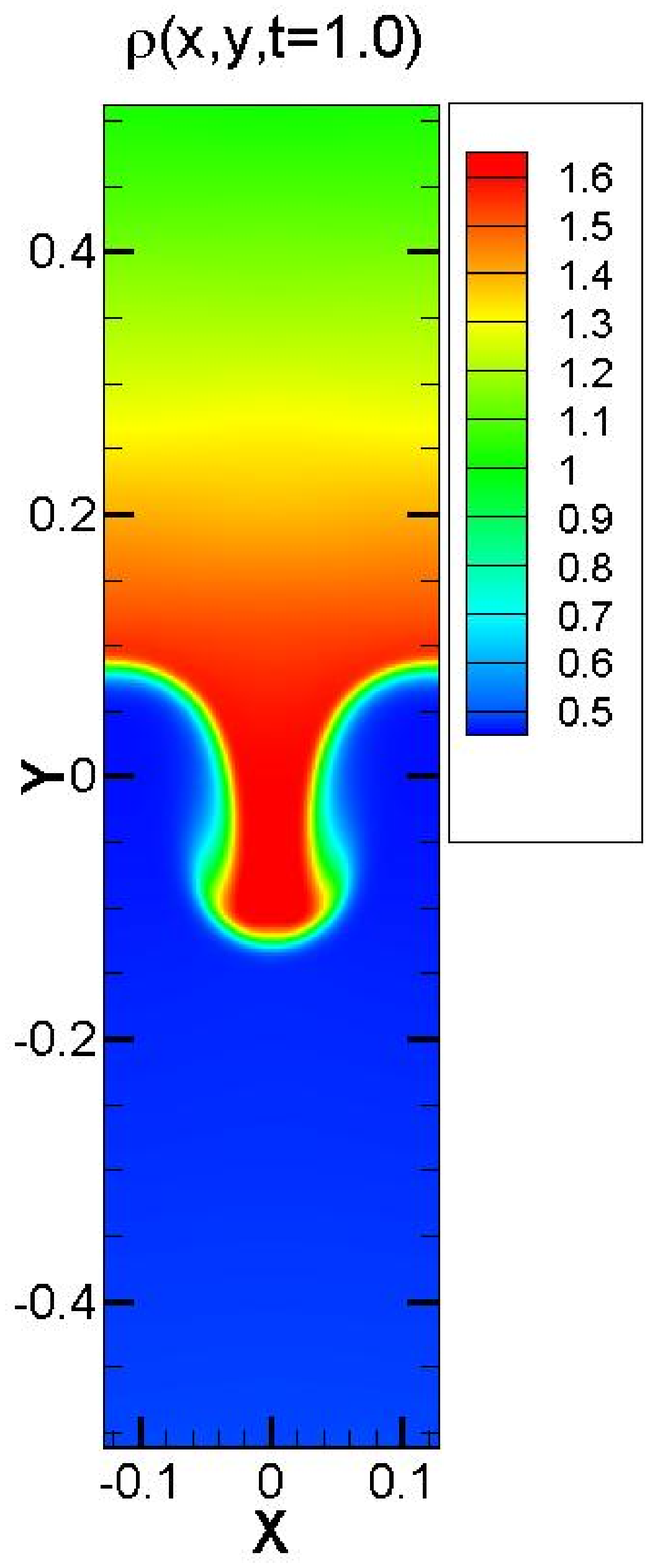,bbllx=22pt,bblly=22pt,bburx=320pt,bbury=742pt,
width=0.165\textwidth,clip=}}\\ \vspace{0.2cm}
{\epsfig{file=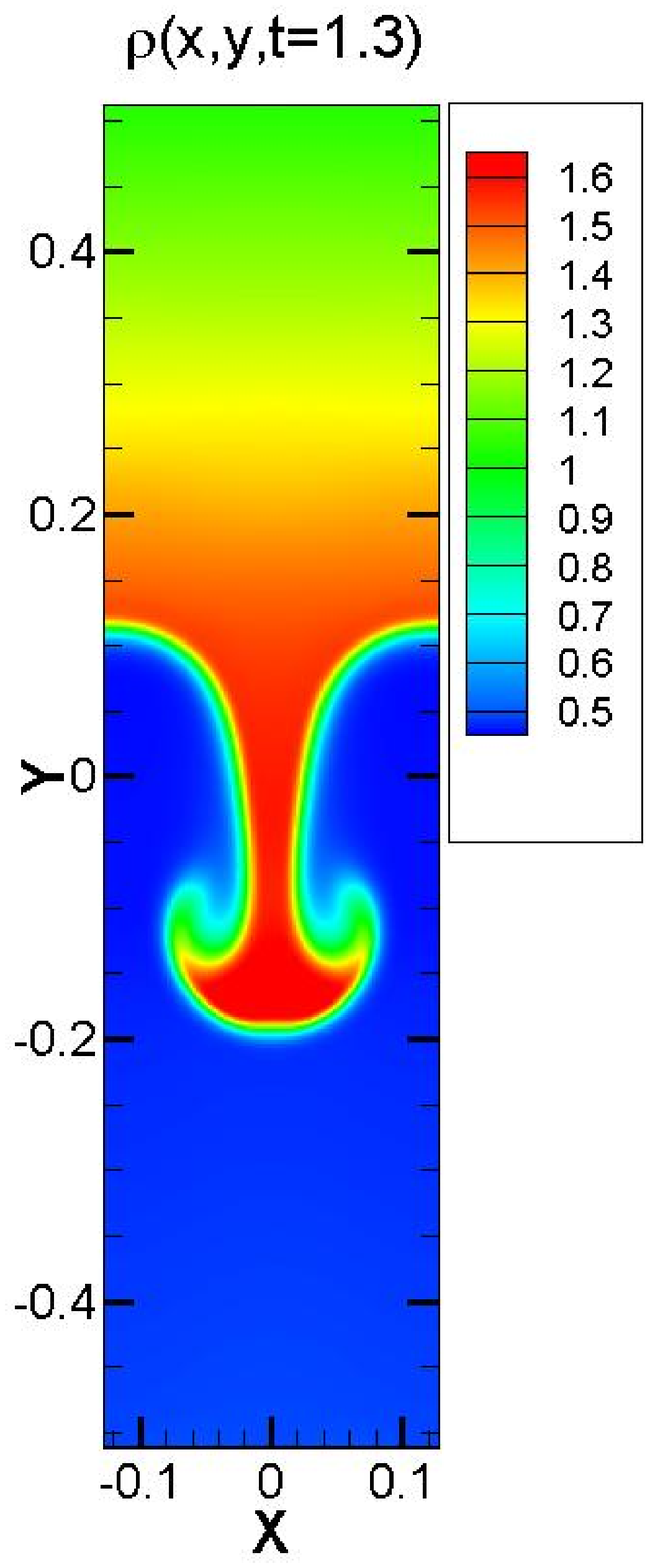,bbllx=22pt,bblly=22pt,bburx=320pt,bbury=742pt,
width=0.165\textwidth,clip=}}\hspace{0.5cm}
{\epsfig{file=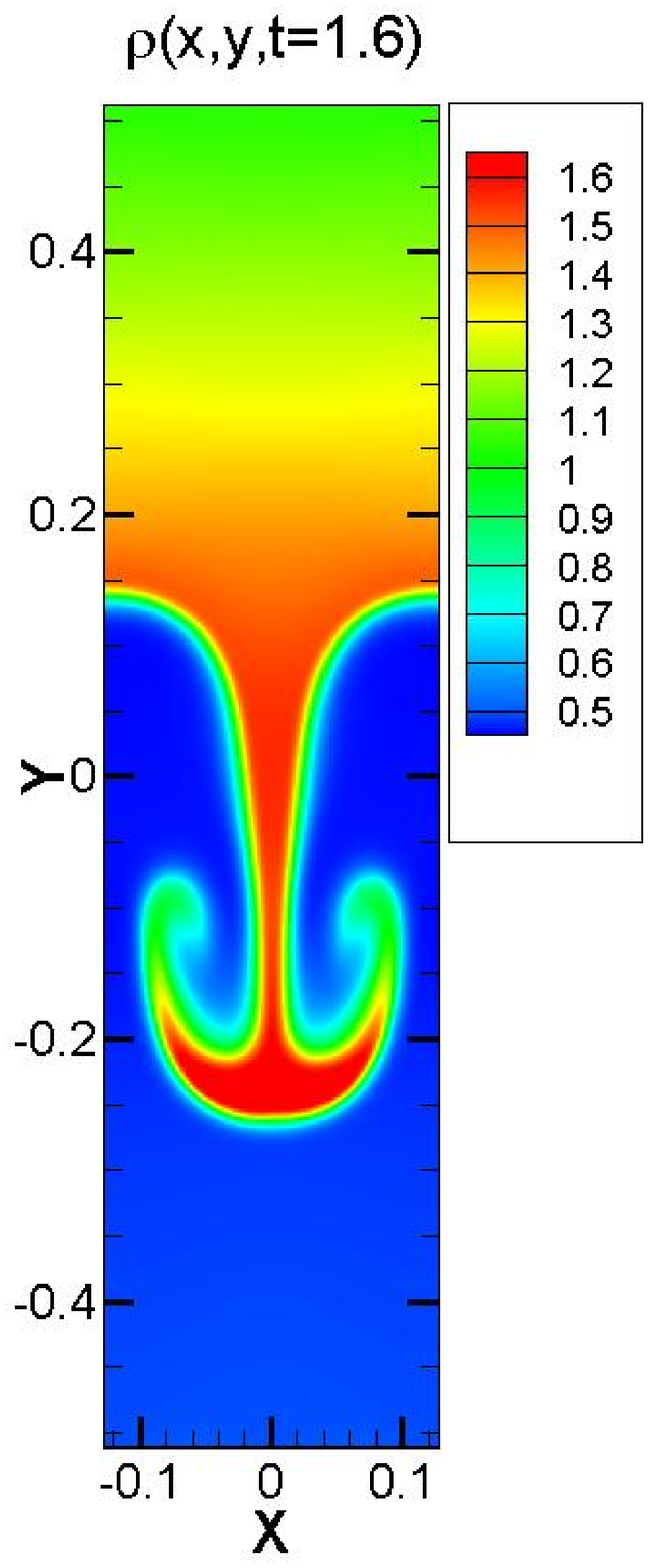,bbllx=22pt,bblly=22pt,bburx=320pt,bbury=742pt,
width=0.165\textwidth,clip=}}\hspace{0.5cm}
{\epsfig{file=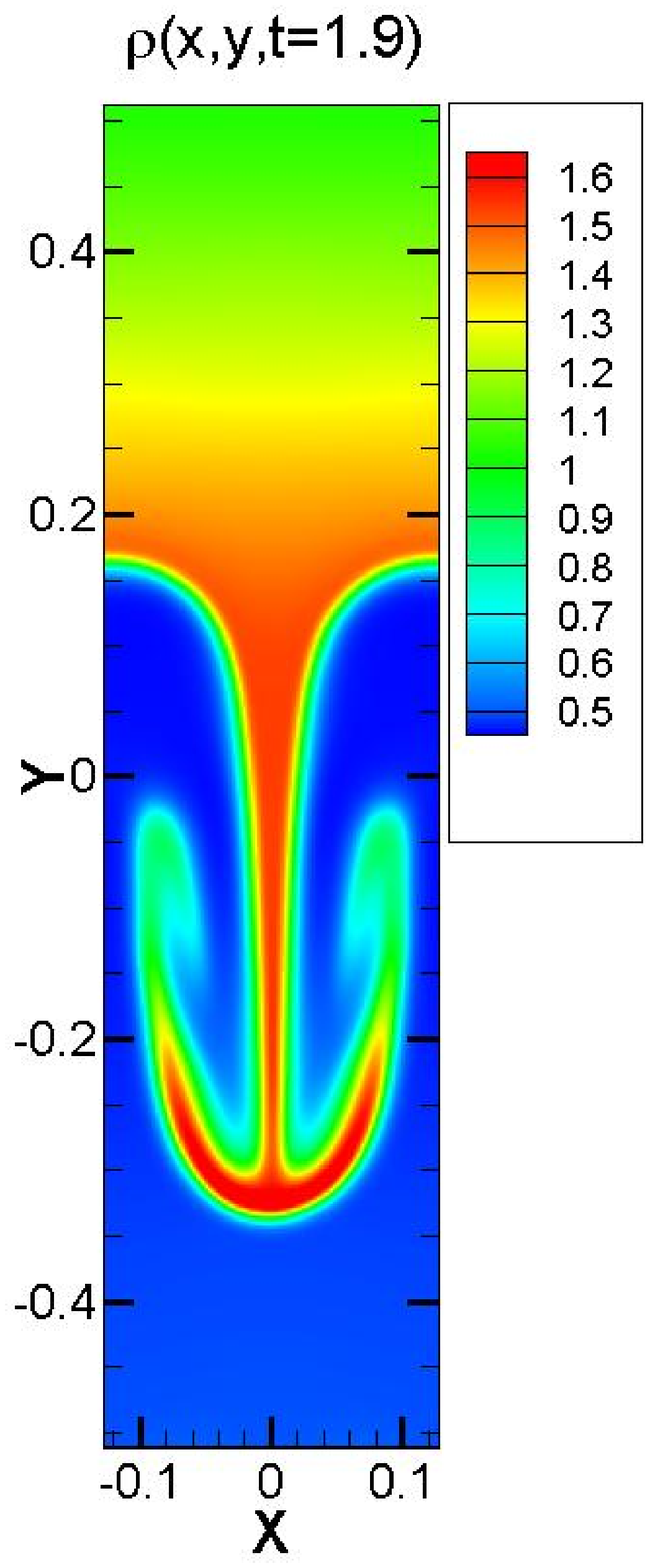,bbllx=22pt,bblly=22pt,bburx=320pt,bbury=742pt,
width=0.165\textwidth,clip=}}\hspace{0.5cm}
{\epsfig{file=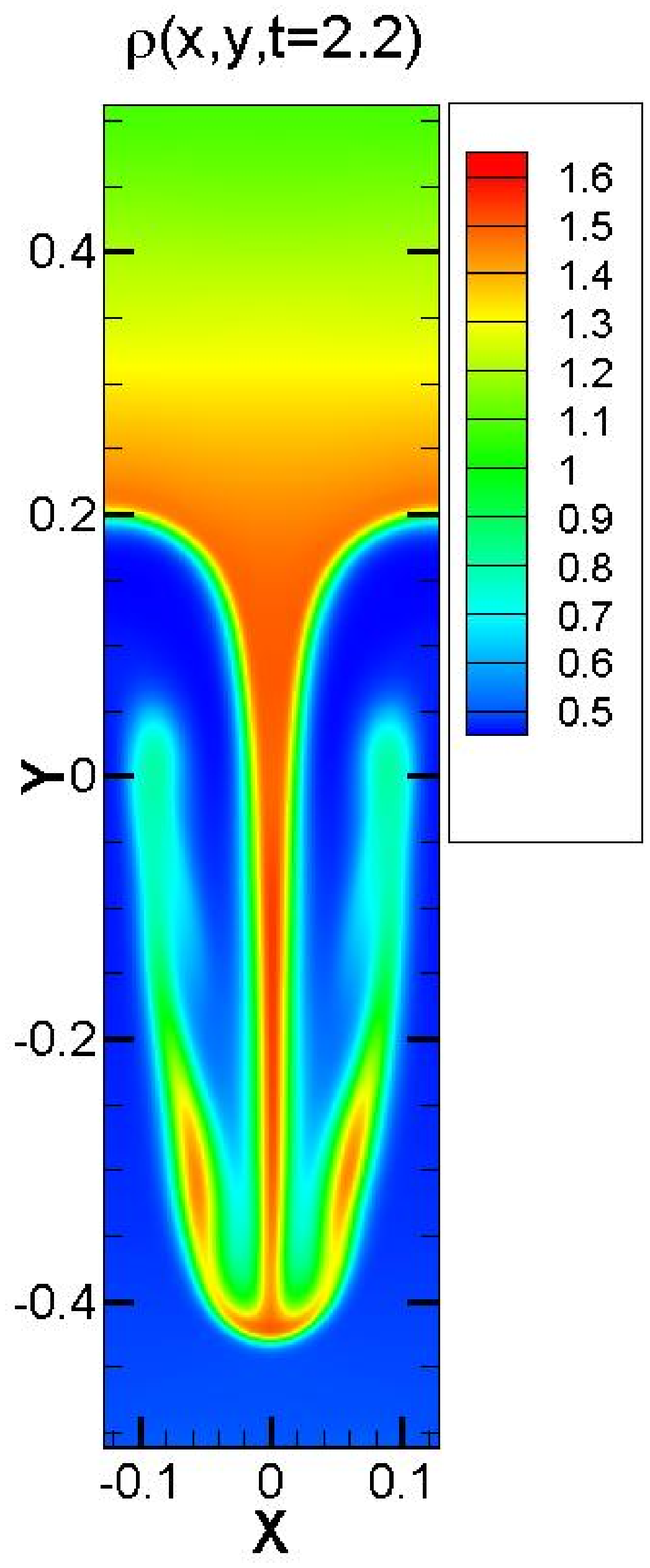,bbllx=22pt,bblly=22pt,bburx=320pt,bbury=742pt,
width=0.165\textwidth,clip=}}
}
\caption{(Color online) Density evolutions in the RTI simulated by the DBM at various times. The same results can also be obtained from the Navier-Stokes model. The larger the density, the stronger the inertial effects, the more difficult to change the velocity. Consequently, the upward perturbation grows into a shape similar to bubble, while the downward perturbation grows into a shape similar to spike.}
\label{FIG1}
\end{figure*}

Here we study both the hydrodynamic and thermodynamic behaviours of the single component compressible RTI.
In our simulations, a grid size of $256\times 1024$ is adopted. The other parameters are $n=3$, $c=1.3$, $\eta_0=15$, $\Delta x=\Delta y=0.001$, $\Delta t=2\times 10^{-5}$, $P_0=1.0$, $a_x=0.0$, $a_y=-g=-1.0$, $\tau=5\times 10^{-5}$, $T_u=1.0$ and $At=0.6$.
Figure \ref{FIG1} displays the density evolutions of RTI. We observe that, at the beginning, thermal diffusion smoothes the discontinuous initial density interface, then a transition layer with finite thickness appears and the local effective Atwood number decreases.
The amplitude of perturbation exponentially grows and the initial configuration remains cosine type until $t=0.6$.
After that, the RTI enters the nonlinear stage, highlighted by the outstanding spike and the appearance of Kelvin-Helmholtz instability due to the difference of the tangential velocity at the interface.

%%%%%%%%%%%%%%%%%%%%%%%%%%%%%%%%%%%%%%%%%%%%%%%%%%%%%%%%%%%%%%%%%%%%%%%%%%%%%%%%
%%%%%%%%%%%%%%%%%%%%%%%%%%%%%%%%%%%%%%%%%%%%%%%%%%%%%%%%%%%%%%%%%%%%%%%%%%%%%%%%%

Owing to the effects of gravity, the fingers of lighter fluid continuously penetrate into the heavier fluid, while the heavier fluid falls into the lighter one with a rolling-up process, resulted in the increase in the mixing layer amplitude, and forming a pair of secondary vortices which appear at the tails of the roll-ups, just like ``mushroom" shape.
The bubble rises due to the release of the compressive energy from the lighter fluid to the heavier one.
In the later stage, since the effects of the viscosity and thermal diffusion, the tails on both vortices
gradually become less sharp and long-narrow. The simulation results are qualitatively consistent with that of the experiments \cite{Lewis,Emmons}, reflecting the basic characteristics of the real physical process.

\begin{figure*}[!ht]
\center {
\epsfig{file=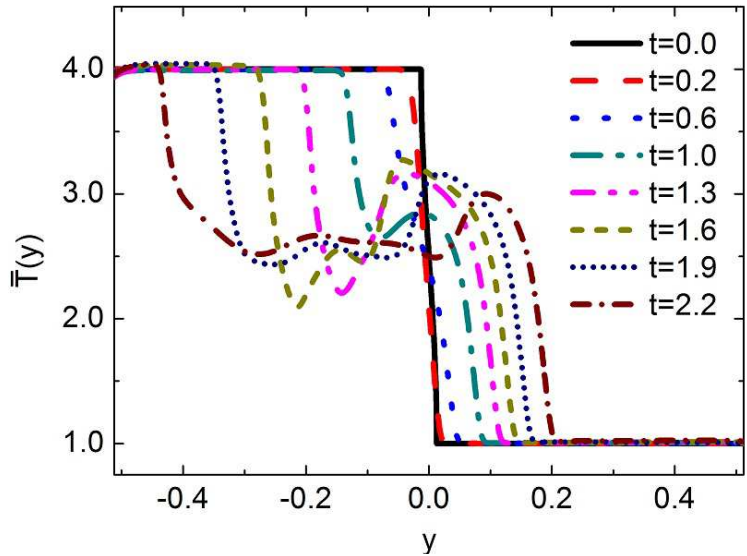,bbllx=53pt,bblly=38pt,bburx=515pt,bbury=380pt,
width=0.6\textwidth,clip=}}
\caption{(Color online) Temperature profiles averaged in the $x$-direction versus $y$ coordinate at different times. The mixing layer becomes wider with time. The perturbation depth in the upper part is smaller than that in the lower part.}
\label{FIG2}
\end{figure*}
%%%%%%%%%%%%%%%%%%%%%%%%%%%%%%%%%%%%%%%%%%%%%%%%%%%%%%%%%%%%%%%%%%%%%%%%%%%%%%%%%

To quantitatively describe the characteristics of the mixing layer, we plot the averaged temperature profile $\overline{T}(y)$ against the $y$ axis at $t=0.0$, $0.2$, $0.6$, $1.0$, $1.3$, $1.6$, $1.9$, and $2.2$ in Fig. \ref{FIG2}. $\overline{T}(y)$ is defined as
\begin{equation}\label{eq7}
\overline{T}(y)=\dfrac{1}{L}\int_LT(x,y)dx.
\end{equation}
It varies from being discontinuous to being irregular that shows the thickness of the mixing layer and the amplitude of the temperature oscillation increases with time. The zig-zags in the profiles indicate the heat conduction of fluids from the high temperature area to the low temperature region and the irregularity in the mixing layer.

\subsection{TNE characterizations of RTI and corresponding interface-tracking technique}

Through the DBM, we can study not only the HNE behaviours, but also the TNE effects of RTI. The TNE effects can be interpreted as the manifestations of molecular thermo-fluctuations relative to the macroscopic flow velocity $\mathbf{u}$ and can therefore help in gaining a better understanding of the kinetic effects on the
onset and development of the RTI.

\begin{figure*}[!ht]
\center {
{\epsfig{file=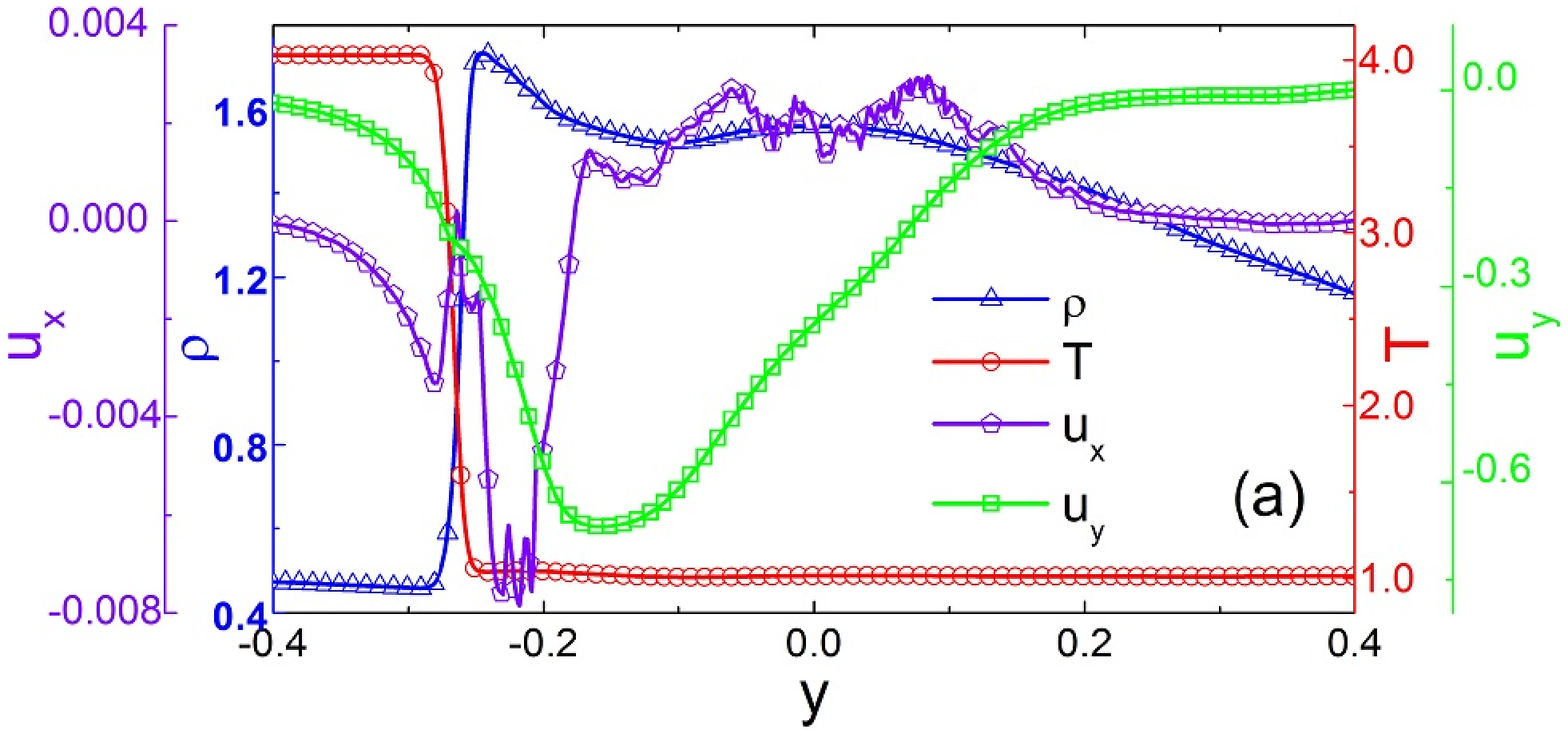,bbllx=23pt,bblly=25pt,bburx=590pt,bbury=290pt,
height=0.28\textwidth,width=0.5\textwidth,clip=}}\\
{\epsfig{file=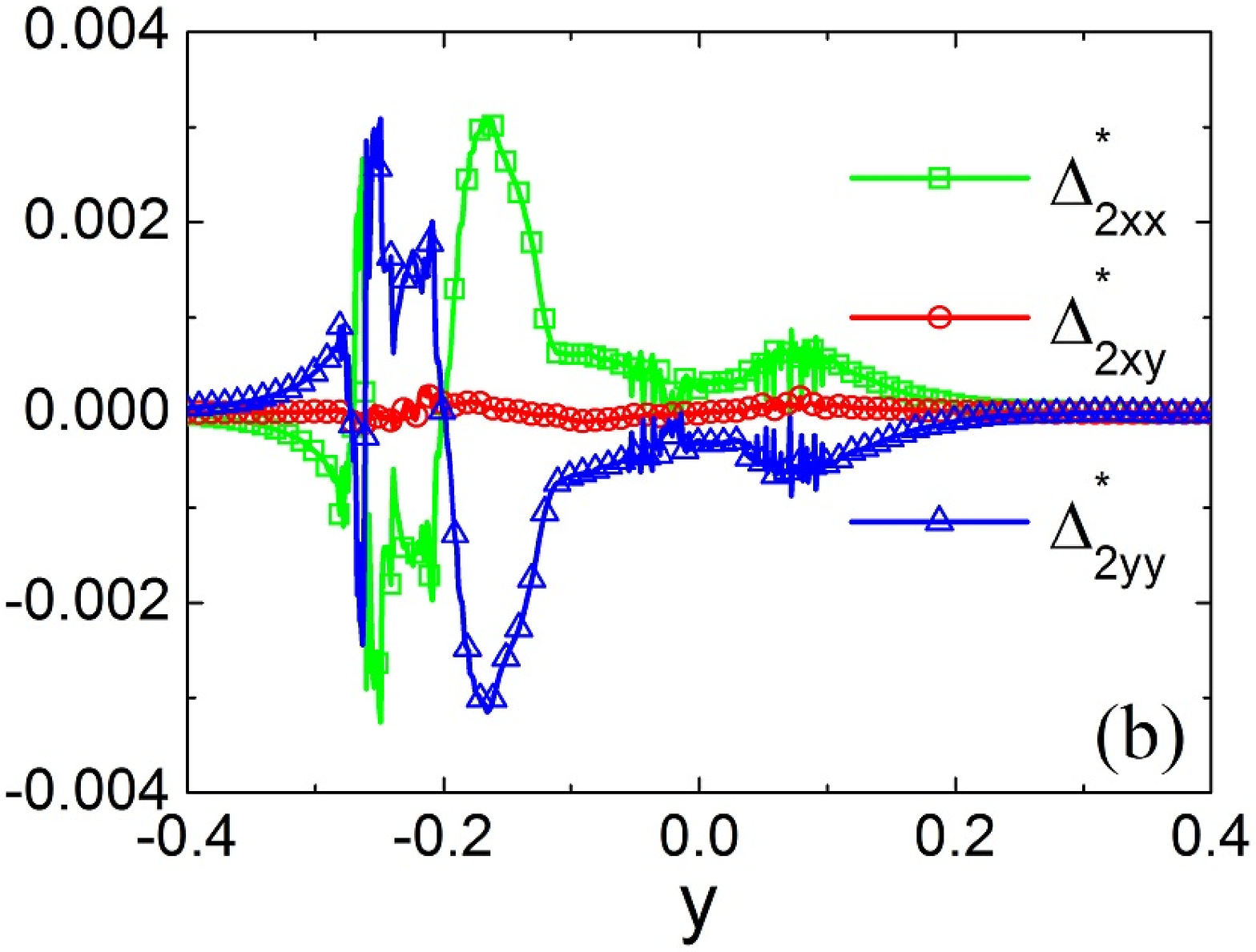,bbllx=50pt,bblly=25pt,bburx=530pt,bbury=390pt,
height=0.25\textwidth,width=0.35\textwidth,clip=}}\hspace{0.5cm}
{\epsfig{file=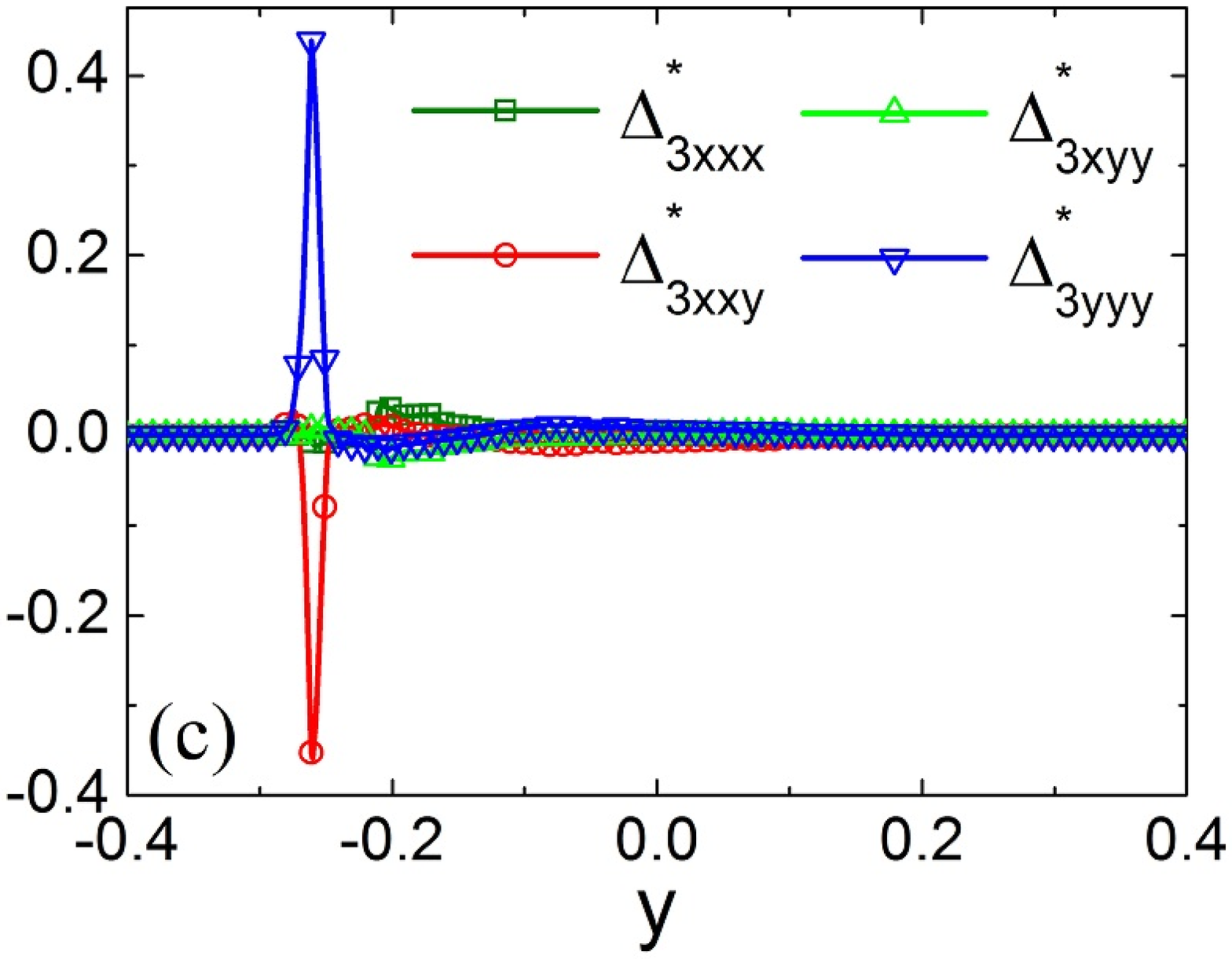,bbllx=50pt,bblly=25pt,bburx=530pt,bbury=390pt,
height=0.25\textwidth,width=0.35\textwidth,clip=}}\\
{\epsfig{file=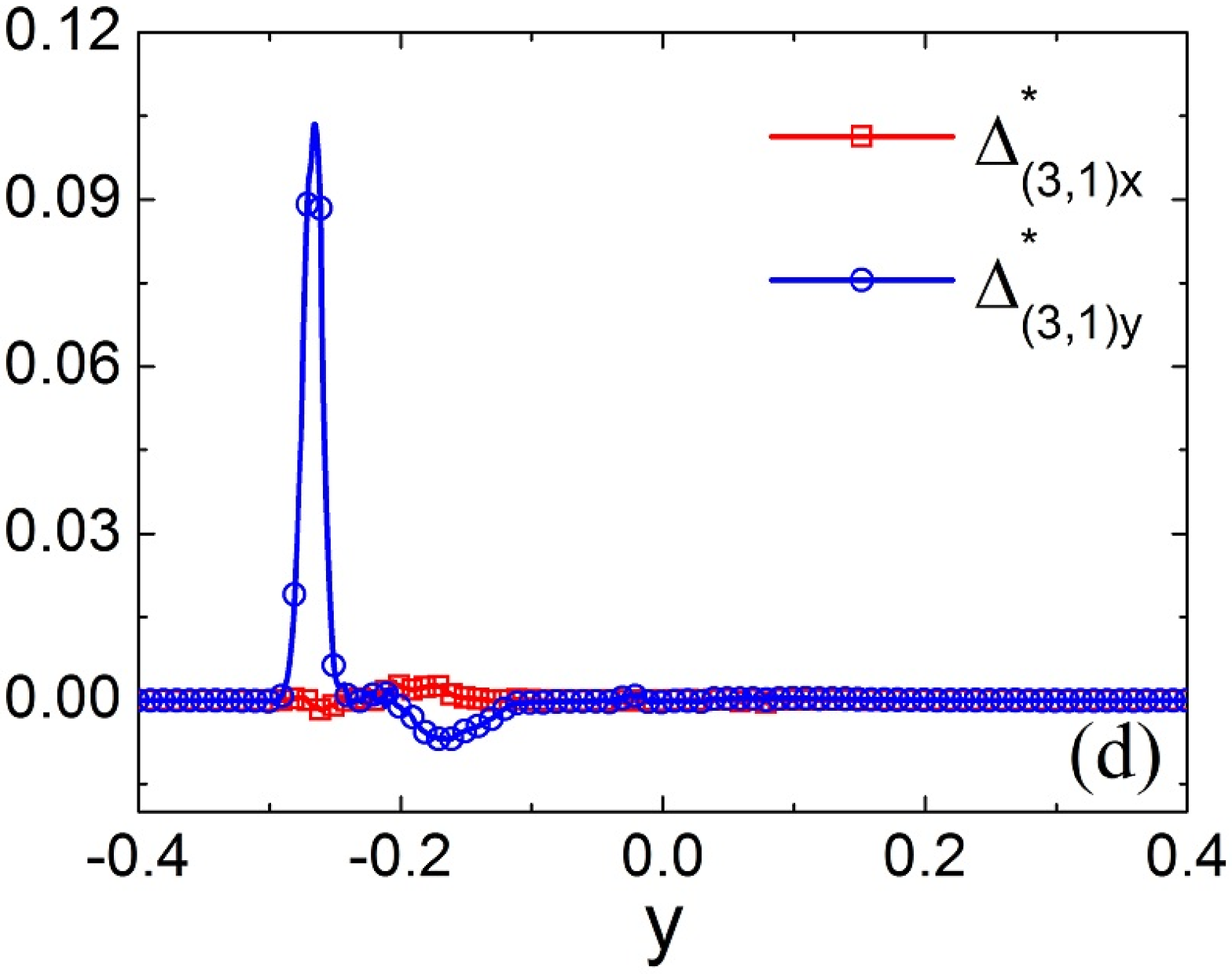,bbllx=50pt,bblly=25pt,bburx=530pt,bbury=390pt,
height=0.25\textwidth,width=0.35\textwidth,clip=}}\hspace{0.5cm}
{\epsfig{file=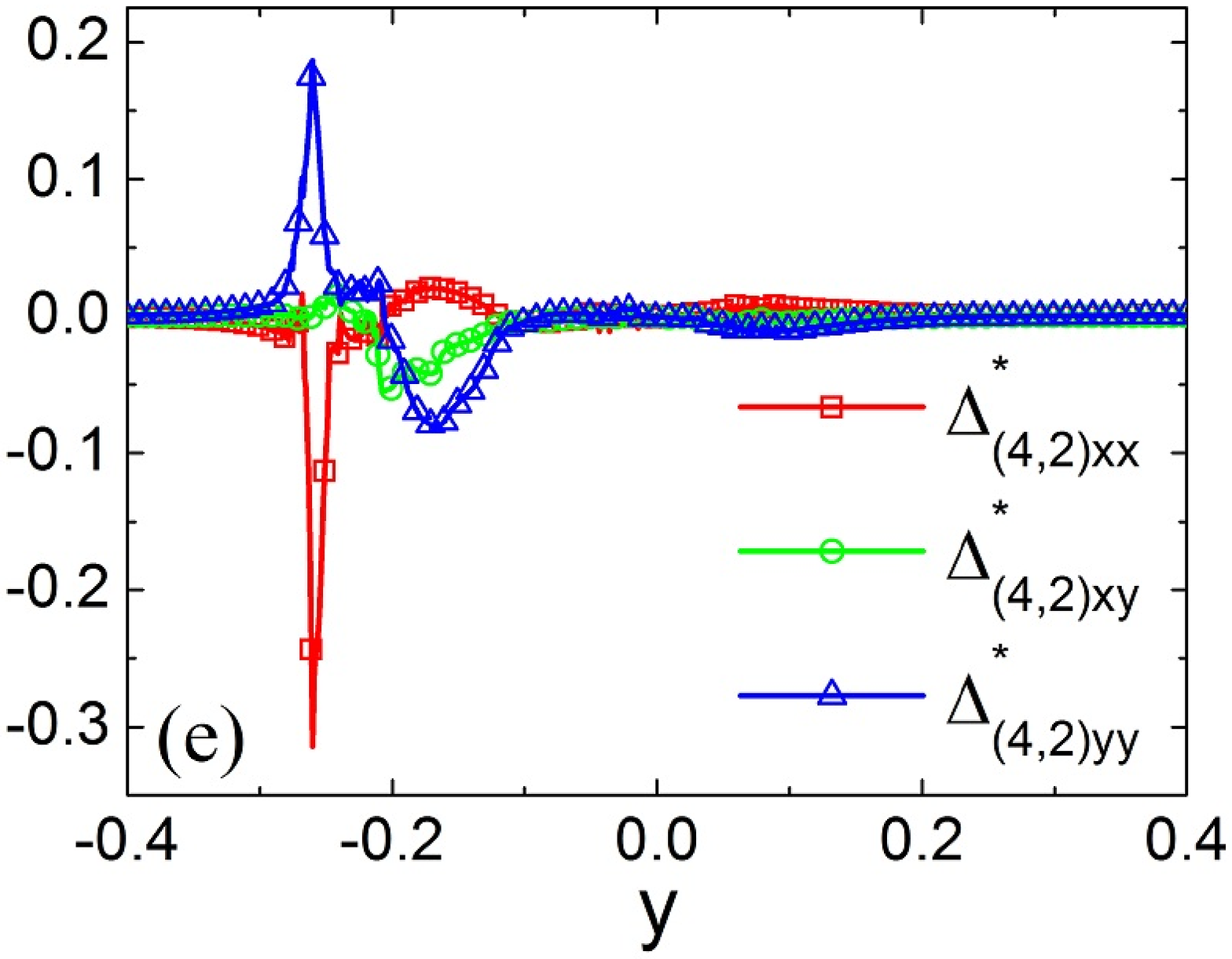,bbllx=50pt,bblly=25pt,bburx=530pt,bbury=390pt,
height=0.25\textwidth,width=0.35\textwidth,clip=}}
}
\caption{(Color online) Profiles of macroscopic quantities and non-equilibrium effects along the central line $x=N_{x}/2$ at $t=1.6$. The gradients of macroscopic quantities work as driving forces of the TNE effects. The TNE quantities show specific TNE status of the system. The most pronounced TNE effects occur around the interface where the macroscopic quantities have large gradients. Because the gravity is in the $y$-direction, the ``flux'' in the $y$-direction is more pronounced. The non-zero value of $\Delta^{*}_{2xy}$ is a typical TNE quantity.}
\label{FIG3}
\end{figure*}
%%%%%%%%%%%%%%%%%%%%%%%%%%%%%%%%%%%%%%%%%%%%%%%%%%%%%%%%%%%%%%%%%%%%%%%%%%%%%%%%%

In Fig. \ref{FIG3}, we illustrate the profiles of macroscopic and TNE quantities along the central line $x=N_{x}/2$ at $t=1.6$. From Fig. \ref{FIG3} (a) one can observe that all the macroscopic quantities show complex behaviours around the interface. The temperature and density profiles show the largest and second largest gradients at this time, respectively. Non-equilibrium effects are mostly pronounced around the interface, where the gradients of macroscopic quantities are particularly strong. Figure \ref{FIG3} (b) shows that the internal kinetic energies in the $x$ and $y$ degrees of freedom deviate from their equilibrium value oppositely in the same amplitude around the interface. Each one of $\Delta^{*}_{2xx}$ and $\Delta^{*}_{2yy}$ deviates from zero oppositely in front of and behind the interface. $\Delta^{*}_{2xy}$, which is zero at equilibrium, shows small but finite values around the interface, which is a typical TNE effect.
From Figs. \ref{FIG3} (c)-(e) we can also appreciate that $\Delta^{*}_{3yyy}$, $\Delta^{*}_{3xxy}$, $\Delta^{*}_{(3,1)y}$, $\Delta^{*}_{(4,2)xx}$, $\Delta^{*}_{(4,2)yy}$ own peaks at the interface. Around the interface, $\Delta^{*}_{3yyy} > 0$,  $\Delta^{*}_{3xxy} < 0$. The value of $\Delta^{*}_{3yyy} + \Delta^{*}_{3xxy} $ in Fig. \ref{FIG3} (c) can be read from the curve of $\Delta^{*}_{(3,1)y}$ in Fig. \ref{FIG3} (d). The positive peak of $\Delta^{*}_{(3,1)y}$ signals an upward internal kinetic energy flux.
This is plausible,  because heat transfers from higher to the lower temperature regions.
In Fig. \ref{FIG3} (e), besides the non-zero value of $\Delta^{*}_{(4,2)xy}$, $\Delta^{*}_{(4,2)xx}$ shows
a larger amplitude than $\Delta^{*}_{(4,2)yy}$ at this time.

\begin{figure}[!ht]
\center {
\epsfig{file=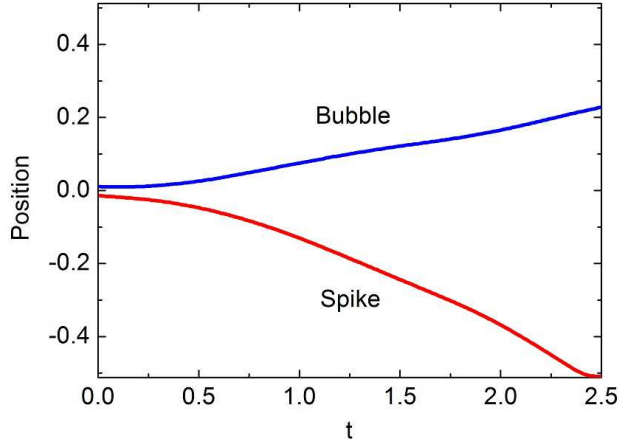,bbllx=48pt,bblly=35pt,bburx=530pt,bbury=387pt,
width=0.5\textwidth,clip=}}
\caption{(Color online) Positions of the bubble and spike versus time. The velocity of the bubble is smaller than that of the spike. The lighter fluid is ``softer'' and thus it is easier for the spike to pass through and grow.}
\label{FIG4}
\end{figure}

\begin{figure}[!ht]
\center {
\epsfig{file=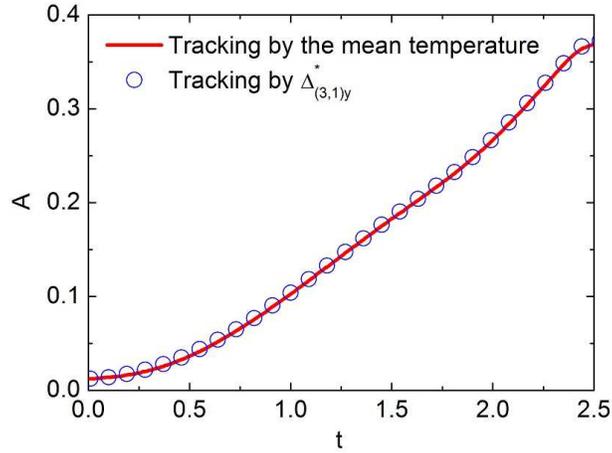,bbllx=50pt,bblly=35pt,bburx=530pt,bbury=390pt,
width=0.5\textwidth,clip=}}
\caption{(Color online) The perturbation amplitudes obtained by two different tracking approaches. The local TNE stength can be used to track interfaces. The good agreement shows that the two approaches validate each other.}
\label{FIG5}
\end{figure}

\begin{figure}[!ht]
\center {
\epsfig{file=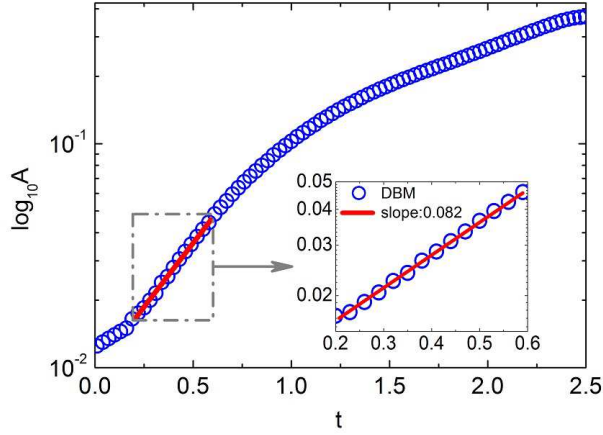,bbllx=36pt,bblly=32pt,bburx=530pt,bbury=380pt,
width=0.5\textwidth,clip=}}
\caption{(Color online) The growth of the perturbation amplitude by the DBM. The good linear fit is consistent with the linear theory on the RTI. The result shows that the linear theory for incompressible flows works also for a period in the compressible flows. The DBM and the linear theory validate each other.
}
\label{FIG6}
\end{figure}

Usually, the depth of the mixing layer is an important parameter to measure the evolution of RTI. We use the half amplitude to measure the mixing layer by capturing the spike and bubble.
For incompressible RTI, this measurement is readily performed by tracing the constant density.
However, in the compressible case, how to measure the mixing layer remains a thorny problem.
Here we present two independent interface-tracking methods:
(i) tracking the mean temperature of the upper and lower fluids, (ii) tracking the maximum values of TNE characteristic quantities, such as $\Delta^*_{(3,1)y}$. The second method is based on the fact that $\Delta^*_{(3,1)y}$ takes its maximum value at the position of interface along the $y$-direction of the spike and bubble. We can adopt this TNE observable to capture the spike and bubble and obtain the thickness of the mixing layer, see Fig. \ref{FIG4}. The agreement between the results obtained from the above two approaches shows the effectiveness of the two tracking schemes, see Fig. \ref{FIG5}. From Fig. \ref{FIG6}, we find that the flow field is qualitatively consistent: from $t=0.2$ to $t=0.6$, the perturbation amplitude grows exponentially,  with a linear growth rate of about 0.082.

%%%%%%%%%%%%%%%%%%%%%%%%%%%%%%%%%%%%%%%%%%%%%%%%%%%%%%%%%%%%%%%%%%%%%%%%%%%%%%%%%
%%%%%%%%%%%%%%%%%%%%%%%%%%%%%%%%%%%%%%%%%%%%%%%%%%%%%%%%%%%%%%%%%%%%%%%%%%%%%%%%%

%%%%%%%%%%%%%%%%%%%%%%%%%%%%%%%%%%%%%%%%%%%%%%%%%%%%%%%%%%%%%%%%%%%%%%%%%%%%%%%%%
%%%%%%%%%%%%%%%%%%%%%%%%%%%%%%%%%%%%%%%%%%%%%%%%%%%%%%%%%%%%%%%%%%%%%%%%%%%%%%%%%

%%%%%%%%%%%%%%%%%%%%%%%%%%%%%%%%%%%%%%%%%%%%%%%%%%%%%%%%%%%%%%%%%%%%%%%%%%%%%%%%%
%%%%%%%%%%%%%%%%%%%%%%%%%%%%%%%%%%%%%%%%%%%%%%%%%%%%%%%%%%%%%%%%%%%%%%%%%%%%%%%%%

\section{Effects of compressibility on RTI}

According to the inviscid, isentropic Euler equation
\begin{equation}\label{eq8}
\partial_{t}\mathbf{u}+\mathbf{u}\cdot \nabla \mathbf{u}+\frac{1}{\rho }
c_{s}^{2}\nabla \rho =\mathbf{g},
\end{equation}
we can introduce a non-dimensional number $H_{1}$ as below:
\begin{equation}\label{eq9}
\dfrac{\left\vert \mathbf{g}\right\vert }{\left\vert \dfrac{\nabla \rho }{\rho
}c_{s}^{2}\right\vert } \sim  \frac{g}{kc_{s}^{2}}=\frac{c_{s}^{-2}}{k/g}%
=H_{1}.
\end{equation}
It is clear that $H_{1}$ can be regarded as the strength of the gravity
relative to the gradient of pressure.  Since $d\rho /dp=c_{s}^{-2}$ describes the compressibility of the flow system, the non-dimensional parameter $H_{1}$ can also be regarded as a relative compressibility. Since the compressibility $c_{s}^{-2}$ is dimensional, it is not suitable for studying the effects of compressibility on RTI. Besides $c_{s}$, both gravity $g$ and the wave number $k$ of the perturbation can influence the increasing rate of RTI. Under such conditions, $H_{1}$ is a good non-dimensional parameter to describe the relative compressibility and is also a good parameter for studying the effects of compressibility on RTI. $H_{1}$ is a variation of speed of sound and consequently a variation of $\gamma $, and also of the stratification width, $1/k$. It can be controlled by adjusting $g$. However, it is totally unrelated  to the viscosity and heat conduction.

Similarly,
\begin{equation}\label{eq10}
H_{2}=\tau \sqrt{gk}=\frac{\tau }{\left( gk\right) ^{-1/2}},
\end{equation}
can be regarded as the ratio of two time scales. Since the relation time $\tau$ is relevant to the viscosity and thermal conductivity, the non-dimensional parameter $H_{2}$ can also be regarded as a relative viscosity or thermal conductivity.

Therefore, we can define $H_1=g/(kc_s^2)$ and $H_2=\tau\sqrt{gk}$ to nondimensionalize the compressibility and the viscosity effects. The dimensionless time is then defined as $t^*=t/\sqrt{2\pi/(kg)}$ and the dimensionless length scale is $2\pi/k$. To study the compressibility $H_1$, we should make sure the viscosity $H_2$ is constant. Furthermore, we can set the initial $\tau=2.0\times 10^{-4}$ and $g=1.0$ to obtain the
prescribed viscosity $H_2$. So we can prescribe $H_1$ by changing $g$.
To ensure $H_2$ is a constant in all simulations, we need to recalculate $\tau$ as derived from $H_2=\tau\sqrt{gk}$ each time. In our numerical simulations, we set the initial Atwood number $At=0.6$, $T_u=1.0$, $P_0=1.0$ uniformly for simplicity. Meanwhile, the mesh is specified by setting $\Delta x=\Delta y=0.001$, and $N_x\times N_y=256\times 1024$. Time step is $\Delta t=2\times10^{-5}$. The other parameters used here are uniformly $n=3$, $c=1.3$ and $\eta_0=15.0$.

%%%%%%%%%%%%%%%%%%%%%%%%%%%%%%%%%%%%%%%%%%%%%%%%%%%%%%%%%%%%%%%%%%%%%%%%%%%%%%%%%
\begin{figure}[!ht]
\center {
\epsfig{file=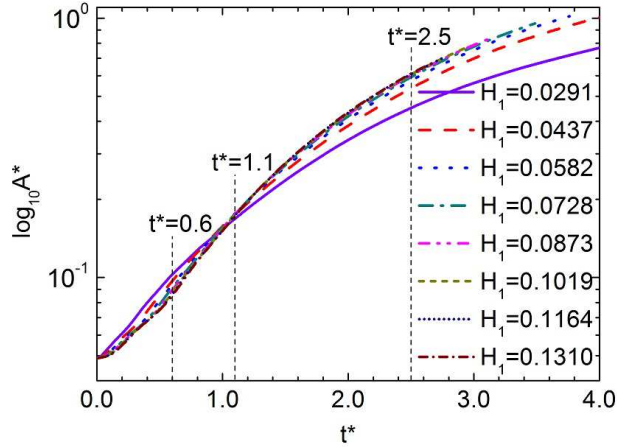,bbllx=47pt,bblly=32pt,bburx=530pt,bbury=383pt,
width=0.5\textwidth,clip=}}
\caption{(Color online) The growth of the dimensionless amplitude $A^*$ with different values of compressibility $H_1$. In the initial stage, the compressibility tends to inhibit the RTI, while in the later stage it tends to strengthen the RTI and the strengthening effect approaches saturation with increasing compressibility.}
\label{FIG7}
\end{figure}
%%%%%%%%%%%%%%%%%%%%%%%%%%%%%%%%%%%%%%%%%%%%%%%%%%%%%%%%%%%%%%%%%%%%%%%%%%%%%%%%%

%%%%%%%%%%%%%%%%%%%%%%%%%%%%%%%%%%%%%%%%%%%%%%%%%%%%%%%%%%%%%%%%%%%%%%%%%%%%%%%%%
\begin{figure}[!ht]
\center {
{\epsfig{file=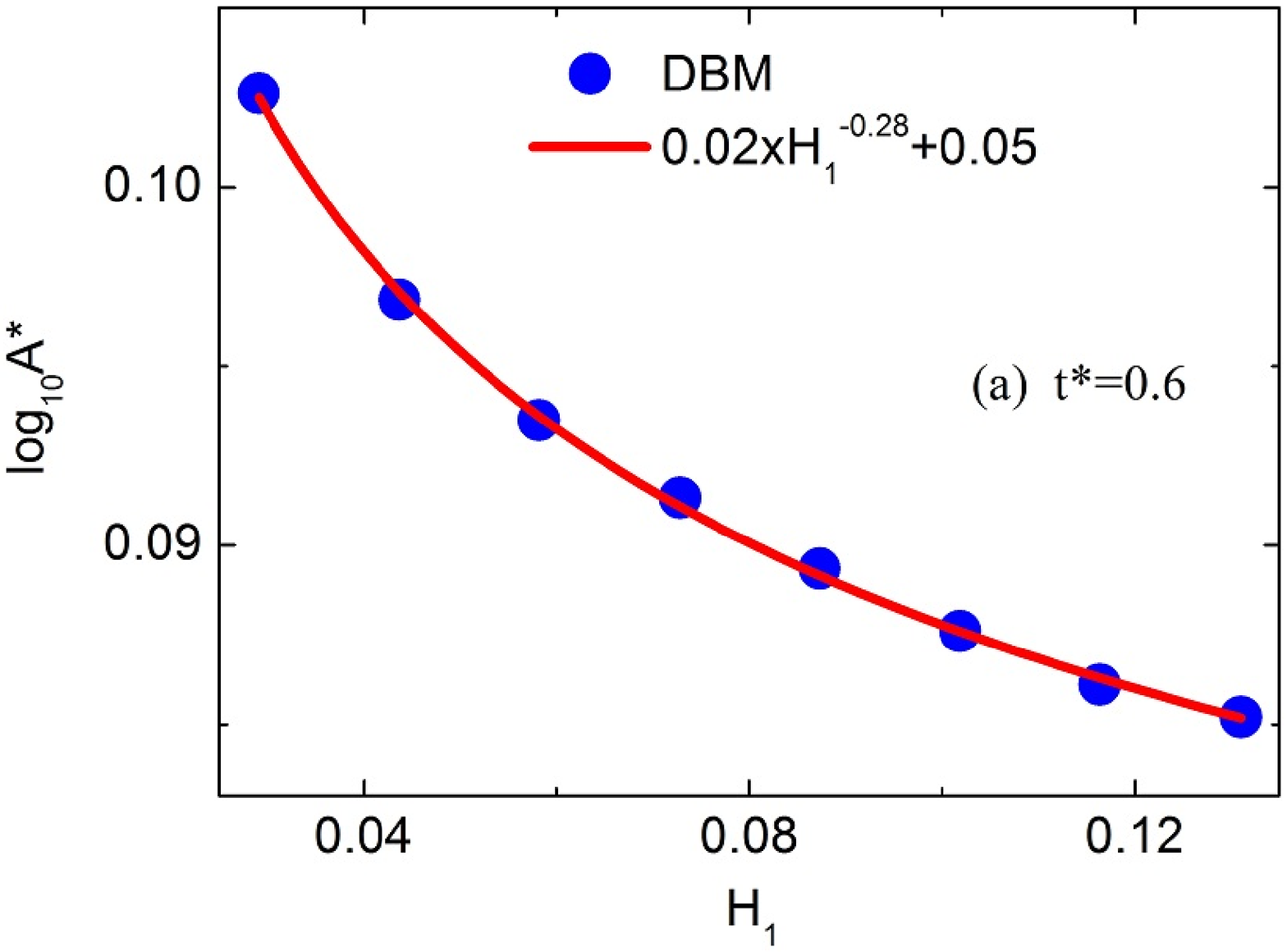,bbllx=32pt,bblly=22pt,bburx=516pt,bbury=380pt,
width=0.43\textwidth,clip=}}
{\epsfig{file=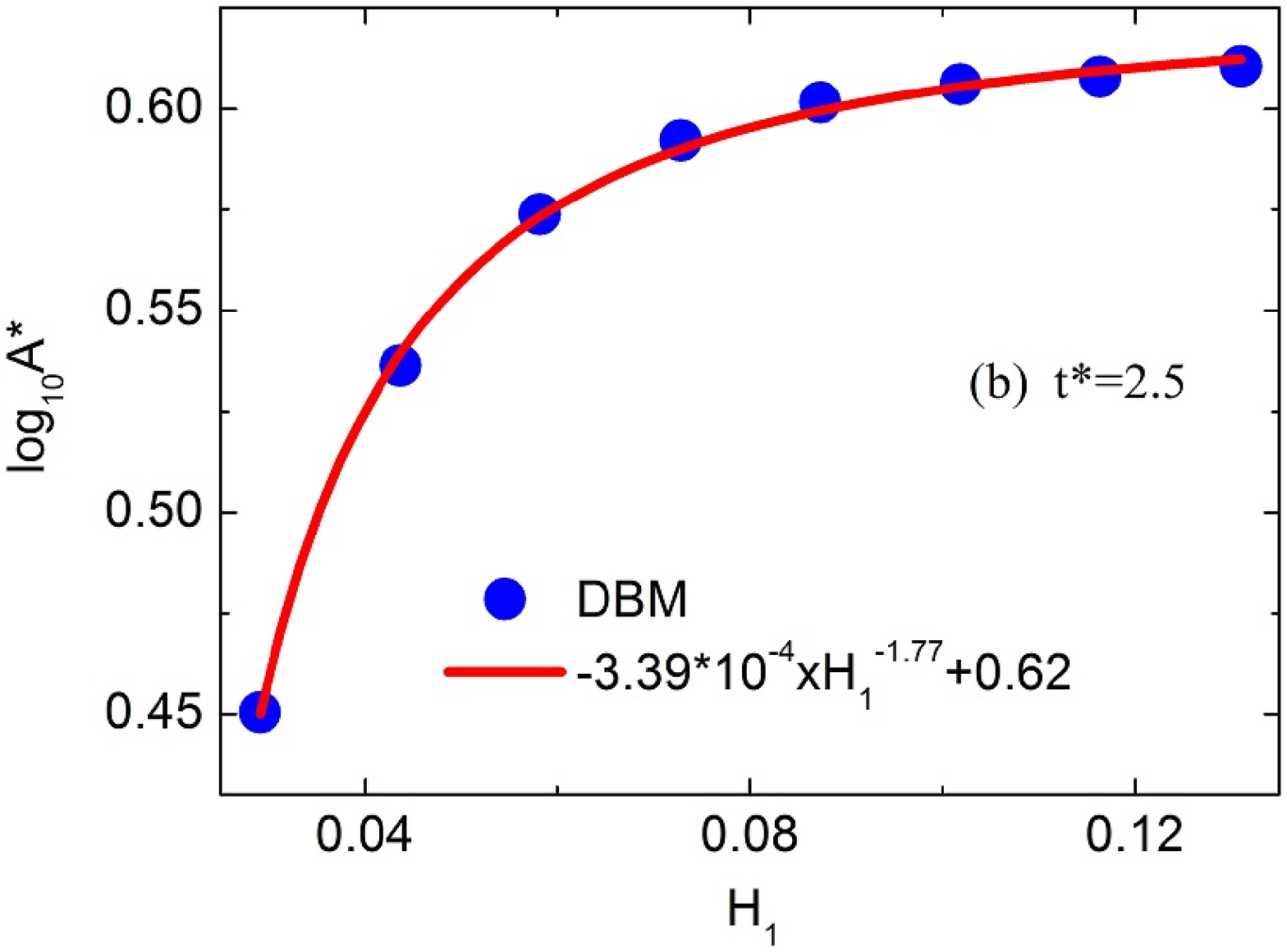,bbllx=32pt,bblly=22pt,bburx=516pt,bbury=380pt,
width=0.43\textwidth,clip=}}
}
\caption{(Color online) The dimensionless amplitude versus
the compressibility at (a) $t^*=0.6$ and (b) $t^*=2.5$.  This figure shows more neatly
the specific compressibility effects at one given time in each of the two stages.}
\label{FIG8}
\end{figure}
%%%%%%%%%%%%%%%%%%%%%%%%%%%%%%%%%%%%%%%%%%%%%%%%%%%%%%%%%%%%%%%%%%%%%%%%%%%%%%%%%

Figure \ref{FIG7} displays the time evolutions of the dimensionless amplitude $A^*$ with
different values of the compressibility. It is found that the effects of compressibility can be roughly divided into two stages: (i) In the first stage, for $t^*<1.1$, compressibility stabilizes the RTI; (ii) In the later stage, for $t^*>1.1$, compressibility accelerates the RTI. Particularly, we show the results at two dimensionless time instants $t^*=0.6$ and $t^*=2.5$, see Fig. \ref{FIG8}, which are fitted by two typical power-law relationships. When the compressibility is sufficiently large, its effects on the growth of RTI become much less evident.

To interpret the above phenomena, we focus on the evolutions of the rates of the system internal and kinetic energies. For the rate of the internal energy, from Eq. (\ref{A8}) (see the Appendix), we have
\begin{equation}\label{eq11}
\rho \dfrac{de}{dt}=-P\nabla\cdot \mathbf{u}+\nabla\cdot(\kappa \nabla T)+\nabla \mathbf{u}:\mathbf{P}^{^{\prime }},
\end{equation}
where $d/dt=\partial/\partial t+\mathbf{u}\cdot \nabla$.
The region that we measure lies in the middle of the system
and its width is half of the system height. This region is large enough to prevent interference of the upper
and lower boundaries with the spike and the bubble, i.e. no heat is supplied to or removed from
the measured region as a result of any interaction with the boundaries.
The heat conduction within the measured region makes no contribution to the increasing rate of internal energy.
Therefore, when considering the rate of internal energy, the second term in the right-hand side of Eq. (\ref{eq11}) can safely be ignored.

We define the rate of the compressive energy, $\dot E_c$, as
\begin{equation}\label{eq12}
\dot E_c=-\int P\nabla\cdot \mathbf{u}dV,
\end{equation}
where $dV$ is the volume element, and the rate of the internal energy due to dissipation or viscosity $\dot E_d$ as
\begin{equation}\label{eq13}
\dot E_d=\int \nabla \mathbf{u}:\mathbf{P}^{^{\prime }} dV.
\end{equation}

For the rate of the kinetic energy, from Eq. (\ref{A8}) (see the Appendix), we have
\begin{equation}\label{eq14}
\rho \dfrac{d\mathbf{u}}{dt}=\rho \mathbf{a}-\nabla P+\nabla\cdot \mathbf{P}^{^{\prime }}.
\end{equation}
This term includes three contributions.
Similarly, we define the rate of the kinetic energy change by gravity, $\dot E_{kg}$, as
\begin{equation}\label{eq15}
\dot E_{kg}=\int \mathbf{u} \cdot (\rho \mathbf{a}) dV,
\end{equation}
the rate of kinetic energy change due to dissipation, $\dot E_{kd}$, as
\begin{equation}\label{eq16}
\dot E_{kd}=\int \mathbf{u} \cdot (\nabla\cdot \mathbf{P}^{^{\prime }}) dV,
\end{equation}
and the rate of kinetic energy change by pressure, $\dot E_{kp}$, as
\begin{equation}\label{eq17}
\dot E_{kp}=-\int \mathbf{u}\cdot\nabla P dV.
\end{equation}

%%%%%%%%%%%%%%%%%%%%%%%%%%%%%%%%%%%%%%%%%%%%%%%%%%%%%%%%%%%%%%%%%%%%%%%%%%%%%%%%%
\begin{figure*}[!ht]
\center {
{\epsfig{file=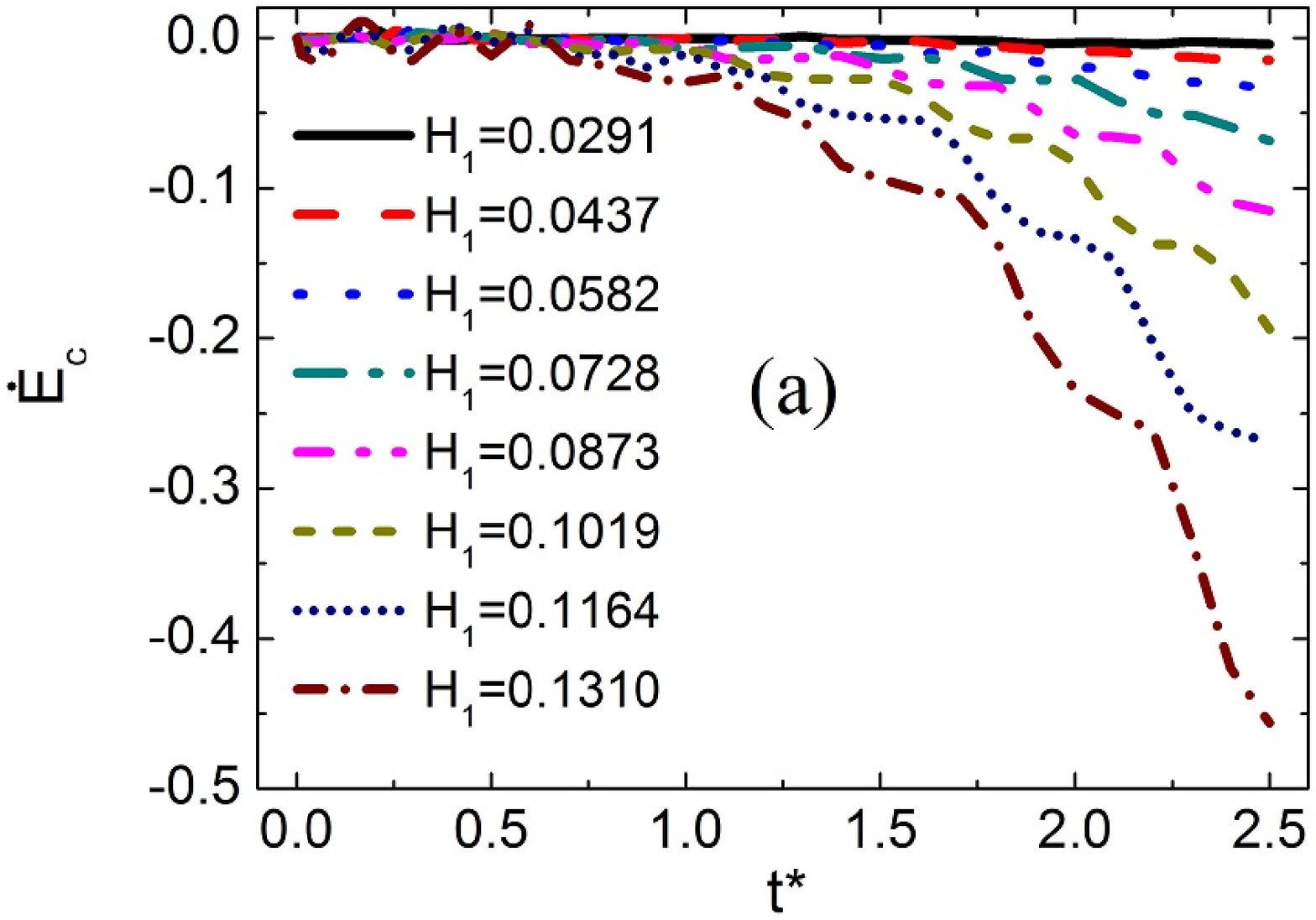,bbllx=26pt,bblly=34pt,bburx=517pt,bbury=382pt,
width=0.45\textwidth,clip=}}\\ \vspace{0.5cm}
{\epsfig{file=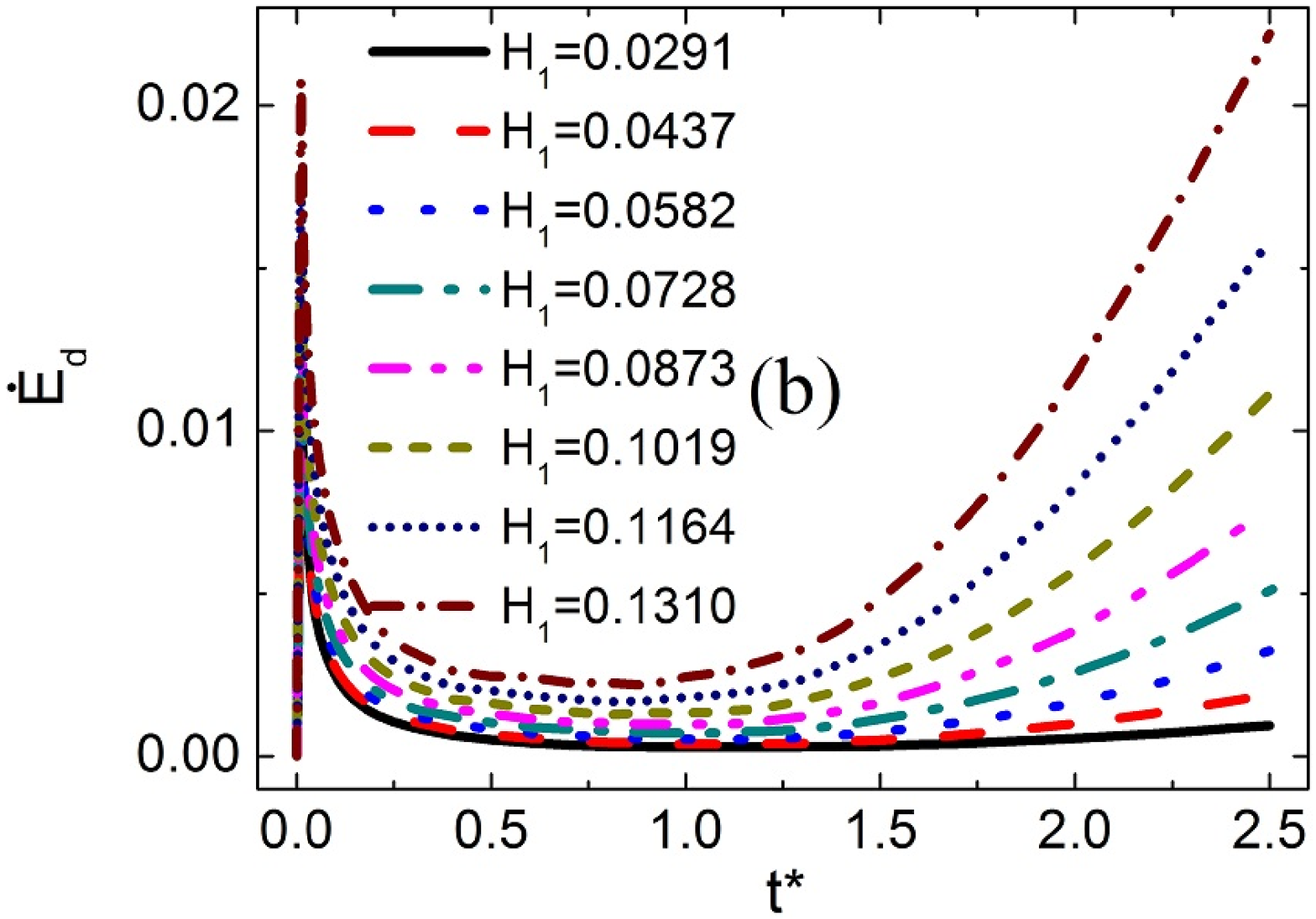,bbllx=26pt,bblly=34pt,bburx=517pt,bbury=382pt,
width=0.45\textwidth,clip=}}\hspace{0.5cm}
{\epsfig{file=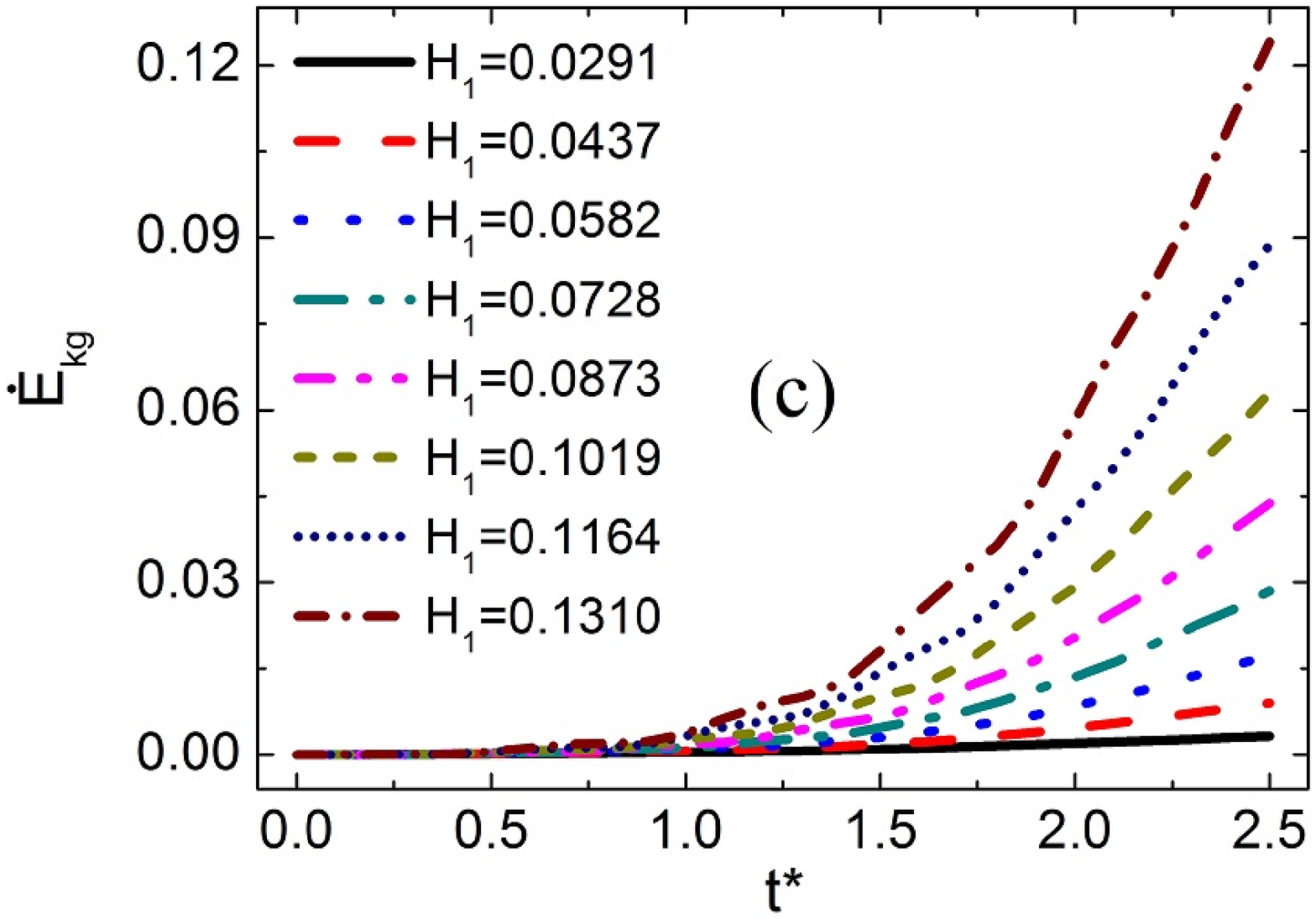,bbllx=26pt,bblly=34pt,bburx=517pt,bbury=382pt,
width=0.45\textwidth,clip=}}\\ \vspace{0.5cm}
{\epsfig{file=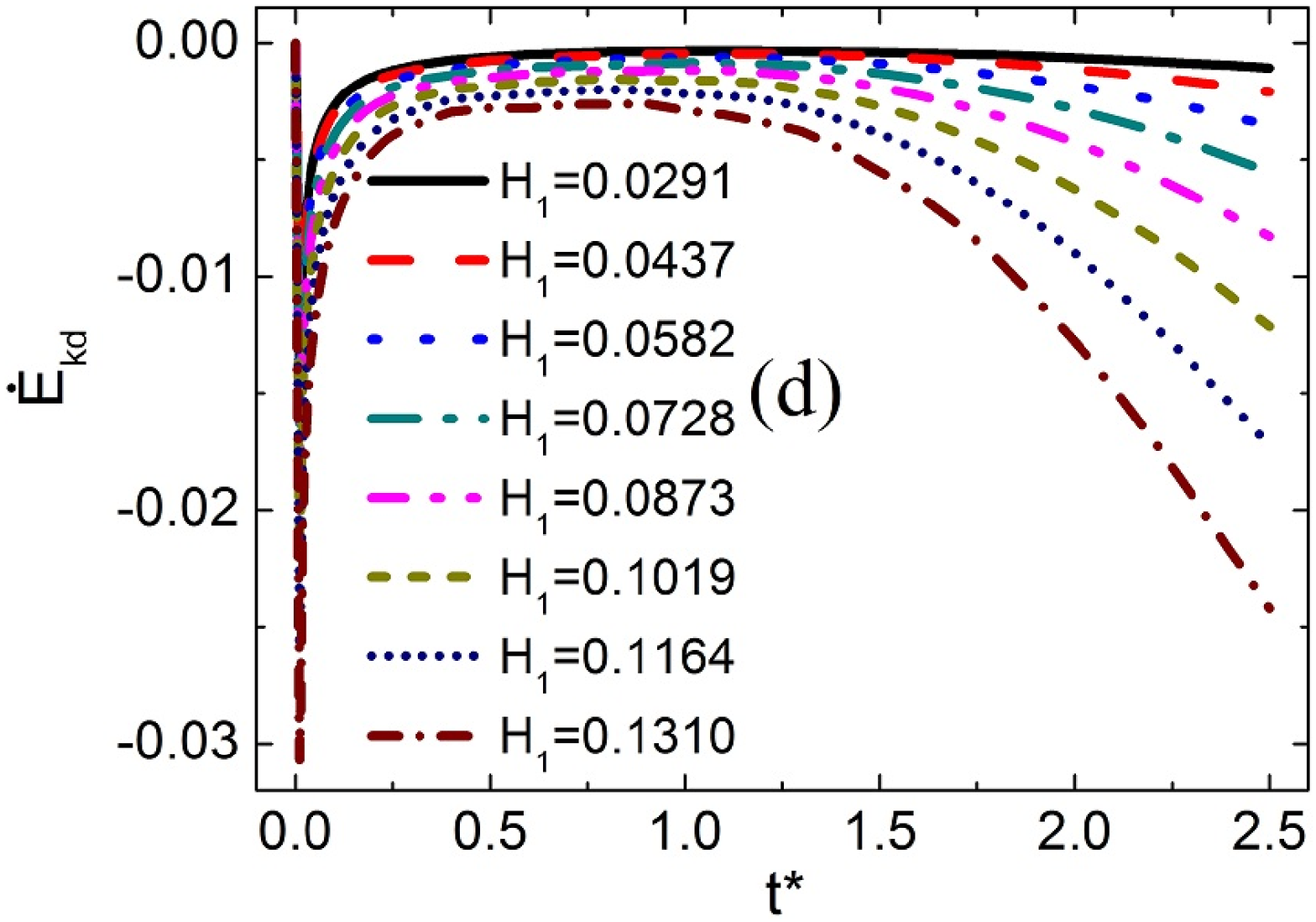,bbllx=26pt,bblly=34pt,bburx=517pt,bbury=382pt,
width=0.45\textwidth,clip=}}\hspace{0.5cm}
{\epsfig{file=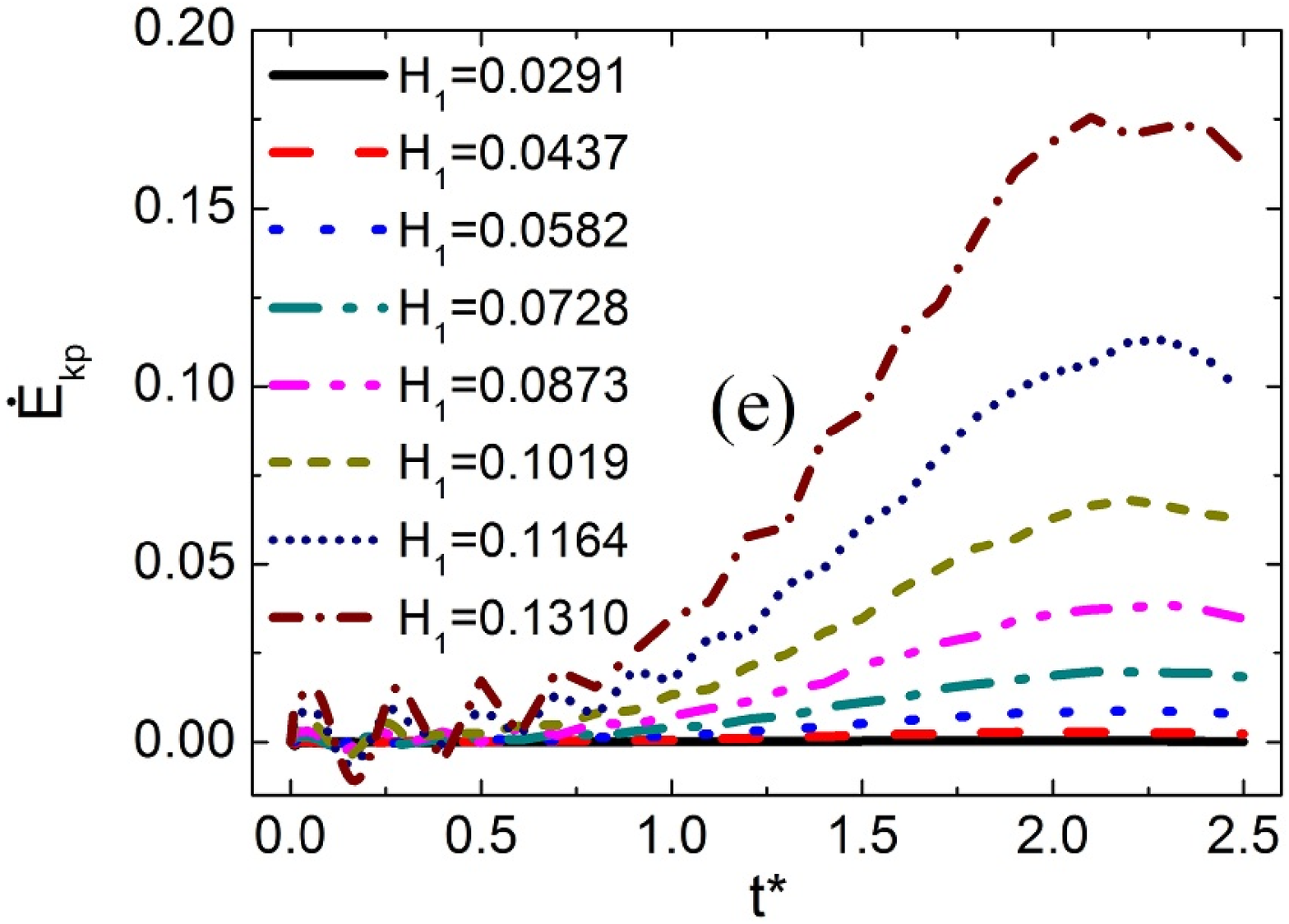,bbllx=26pt,bblly=34pt,bburx=517pt,bbury=389pt,
width=0.45\textwidth,clip=}}
}
\caption{(Color online) The time evolutions of the various energy rates at different values of compressibility $H_1$. The energy rates include the rates of compressive energy $\dot E_c$, internal dissipation energy $\dot E_d$, kinetic energy induced by gravity $\dot E_{kg}$, kinetic energy induced by dissipation $\dot E_{kd}$ and kinetic energy induced by pressure $\dot E_{kp}$. It is clear that a larger compressive energy rate is involved in the later stage. This figure highlights significant differences as compared to the incompressible scenario.}
\label{FIG9}
\end{figure*}
%%%%%%%%%%%%%%%%%%%%%%%%%%%%%%%%%%%%%%%%%%%%%%%%%%%%%%%%%%%%%%%%%%%%%%%%%%%%%%%%%

Figure \ref{FIG9} illustrates the time evolution of the various energy rates, $\dot E_c$, $\dot E_d$, $\dot E_{kg}$, $\dot E_{kd}$ and $\dot E_{kp}$, with different compressibilities.
We first discuss observations for the second stage of RTI evolution.
From Fig. \ref{FIG9} (a) it is clear that with decreasing the compressibility, $\dot E_c$ goes gradually
back to the case of incompressible flows, where $\dot E_c = 0$.
The extent of compressive energy rate $\dot E_c$ increases with time and compressibility.
The rate $\dot E_c$ is negative, which means that the fluid volume expands in time.
Since the fluid has internal dissipation, the rate of the energy dissipation $\dot E_d$
increases in time and with increasing viscosity [see Fig. \ref{FIG9} (b)].
In this study, the compressibility $H_{1}$ is increased by increasing the gravity acceleration $g$, consequently the kinetic energy rate by gravity increases with increasing compressibility $H_{1}$ [see Fig. \ref{FIG9} (c)].
Since the volume expansion rate increases with increasing compressibility and time, the
rate of kinetic energy by dissipation or viscosity $\dot E_{kd}$ show similar behaviour [see Fig. \ref{FIG9} (d)].
It is interesting to observe that the rate of kinetic energy by pressure $\dot E_{kp}$ also increases with the compressibility $H_{1}$ [see Fig. \ref{FIG9} (e)]. This behaviour can be understood by considering that the compressibility $H_{1}$ increase means that gravity acceleration $g$ increases.
Consequently, the pressure gradient, $\nabla P$, becomes larger, thereby enhancing the rate of
kinetic energy by pressure $\dot E_{kp}$ as well [see Eq. (\ref{eq17})].

Now, we come back to the initial stage.
Initially, the flow velocity is low, and so is the Mach number, so that the compressibility
of the fluid is naturally small.
So, the values of $\dot E_c$, $\dot E_{kg}$ and $\dot E_{kp}$ are also small.
The initial state of the system around the interface is thermodynamically unstable.
Without any perturbation, the interface molecules are in a mechanically stable
but thermodynamically unstable state.
In this case, the forces due to the temperature and density gradients
tend to increase the boundary layer and decrease the local Atwood number.
If the interface is initially perturbed,  most  interface molecules, in between the crest and trough, experience
a gradient force with a non-zero horizontal component, which tends to flatten the interface.
It should be pointed out that in our numerical simulations, the initial state is slightly
different from the one given by Eq. (\ref{eq4}), due to the finite lattice spacing and
the free-energy of the numerical initial state is higher than the one given by Eq. (\ref{eq4}).
The system relaxes towards its minimum free energy state and gravitational energy
transforms into kinetic energy.
Due to viscosity, while approaching this minimum free energy state, the local
flow velocity tends to zero, so that the amplitudes of
 $\dot E_{d}$ and $\dot E_{kd}$, after a quick initial rise, decrease significantly in the longer term.
The stage for the initial quick increase is indeed very short.

The process of RTI can also be divided into two stages and be interpreted from the point view of energy transformation: (i) In the initial stage, the compressibility provides a stabilizing effect.
This is mainly because  a higher compressibility $H_{1}$ corresponds to a larger gravity acceleration $g$,
corresponding in turn to a larger local density $\rho$ or pressure $P$, and consequently a higher heat conductivity $\kappa$. The heat conduction tends to decrease the local Atwood number and broaden the interfaces of the density and temperature. (ii) In the later stage, the compressibility has a destabilizing effect, which can be explained as the transformation of the stored compressive energy into kinetic energy.

To provide an estimate of the TNE effects resulting from compressibility, we follow the same idea used in Refs. \cite{XuLin2015PRE} and \cite{XuGan2015}, and define a global
average non-equilibrium intensity or ``TNE strength"
\begin{eqnarray}\label{eq18}
D^{*}=\sqrt{(\bar{\boldsymbol{\Delta}}^*_{2})^2+(\bar{\boldsymbol{\Delta}}^*_{3})^2
+(\bar{\boldsymbol{\Delta}}^*_{3,1})^2+(\bar{\boldsymbol{\Delta}}^*_{4,2})^2},
\end{eqnarray}
where $\bar{\boldsymbol{\Delta}}^*_{m,n}$ is the average absolute value of TNE components.

%%%%%%%%%%%%%%%%%%%%%%%%%%%%%%%%%%%%%%%%%%%%%%%%%%%%%%%%%%%%%%%%%%%%%%%%%%%%%%%%%
\begin{figure}[!ht]
\center {
\epsfig{file=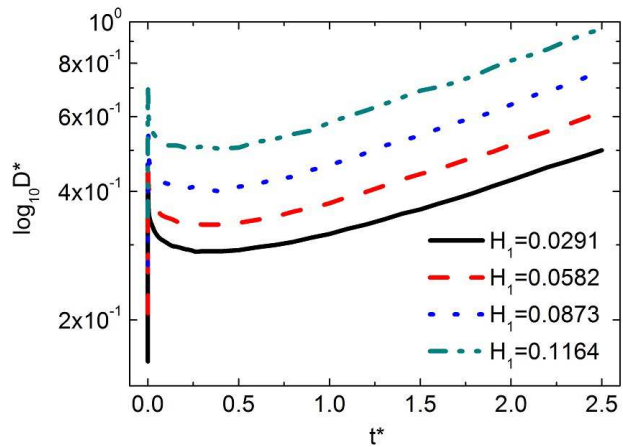,bbllx=23pt,bblly=33pt,bburx=518pt,bbury=391pt,
width=0.5\textwidth,clip=}}
\caption{(Color online) The time evolution of the global average TNE strength with different values of compressibility $H_{1}$. The compressibility first decreases in the first stage and then increases in the later stage the global average TNE strength. The higher the compressibility, the stronger the global average TNE effects. }
\label{FIG10}
\end{figure}
%%%%%%%%%%%%%%%%%%%%%%%%%%%%%%%%%%%%%%%%%%%%%%%%%%%%%%%%%%%%%%%%%%%%%%%%%%%%%%%%%

Figure \ref{FIG10} shows how the compressibility affects the global average $D^{*}$. With increasing compressibility, the global average deviation from thermodynamic equilibrium increases. Since the initial condition is in thermal non-equilibrium state, the system has a tendency to approach the thermodynamic equilibrium state at first. Therefore, in the first stage, $D^{*}$ shows a decreasing trend. In the later stage, $D^{*}$ is found to grow exponentially. This is because TNE effects are tightly coupled to the interface dynamics, which shows an increasing area (coarsening) and morphological complexity.

\begin{figure*}[!ht]
\center {
{\epsfig{file=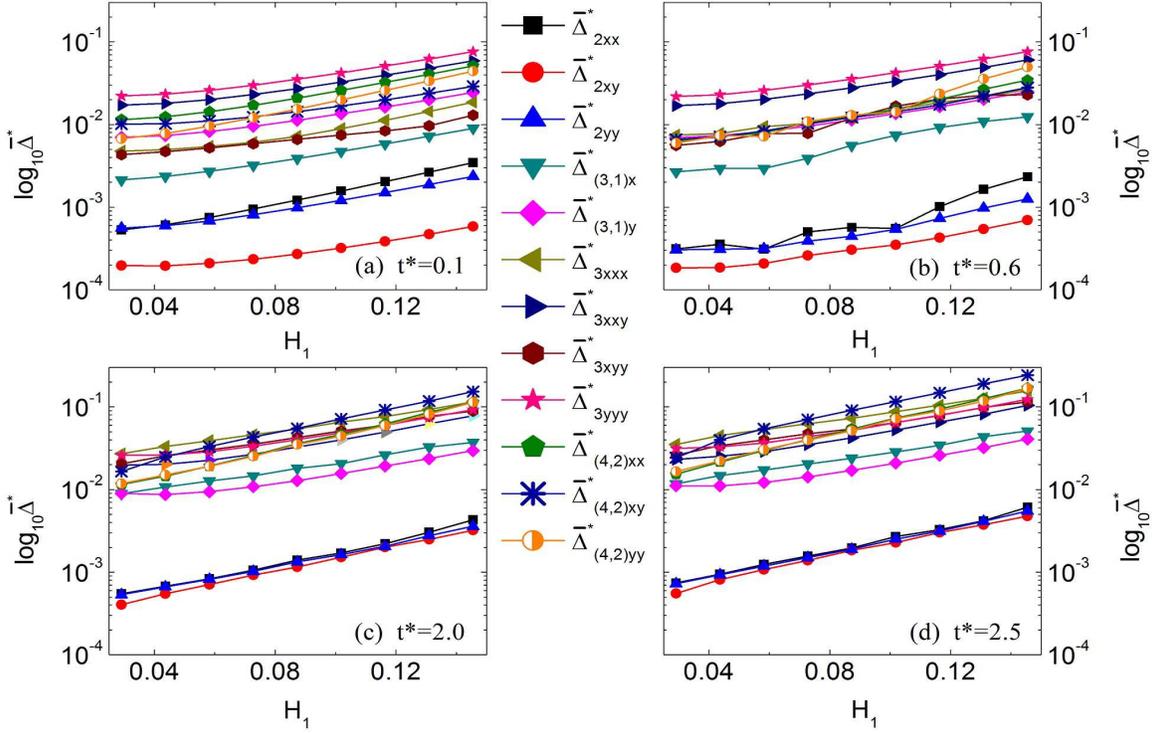,bbllx=19pt,bblly=20pt,bburx=591pt,bbury=400pt,
height=0.6\textwidth,width=0.95\textwidth,clip=}}
}
\caption{(Color online) The average values of various components of TNE quantities versus compressibility at four nondimensional times: $t^*=0.1$, $0.6$, $2.0$ and $2.5$. The figure shows more specific global or mean TNE effects than Fig. \ref{FIG10}. The relative strengths of those dependences on compressibility may change with time.}
\label{FIG11}
\end{figure*}
%%%%%%%%%%%%%%%%%%%%%%%%%%%%%%%%%%%%%%%%%%%%%%%%%%%%%%%%%%%%%%%%%%%%%%%%%%%%%%%%%

Figure \ref{FIG11} shows the average value of each component of TNE observables with changing compressibility at four different times $t^*=0.1$, $0.6$, $2.0$ and $2.5$.
It is found that:
(i) All TNE dynamic modes increase as the compressibility increases, at all times.
Since all the observations have a limited accuracy, a dynamic mode is not visible whenever its amplitude or strength is smaller than a critical value, say $10^{-3}$. Therefore, one can observe that more TNE dynamic modes emerge and stand out with increasing the compressibility.
At later stage, the higher order terms of the TNE dynamic modes, such as $\bar{\Delta}^*_{(4,2)xy}$, play a more important role. A single TNE dynamic mode is not the only decisive factor for the amplitude of the $D^{*}$; (ii) The relative strength among the dynamic modes may change at different stages, for example, $\bar{\Delta}^*_{(3,1)x}$ is less than $\bar{\Delta}^*_{(3,1)y}$ at the first stage, but reverse at the later stage; (iii) The strengths of some dynamic modes are always relatively small, such as $\bar{\Delta}^*_{2xx}$, $\bar{\Delta}^*_{2xy}$ and $\bar{\Delta}^*_{2yy}$. These details are complementary with the above highly synthetic ``TNE strength" $D^{*}$.

\section{Conclusions}

The Rayleigh-Taylor instability in compressible flows is studied via a discrete Boltzmann model.
Besides hydrodynamic behaviour, the thermodynamic
non-equilibrium effects most relevant to the hydrodynamic behaviour have also been studies in much detail, up
to Atwood numbers around $0.9$.
It is found that the process of the Rayleigh-Taylor instability in compressible flows
can be divided into two stages, exhibiting opposite compressibility effects: in the initial stage, compressibility
stabilizes the Rayleigh-Taylor instability, while in the later stage it accelerates it.
The physical reasons are as follows.
A higher compressibility leads to a stronger gravity acceleration, which corresponds to
a higher local pressure and consequently to a higher heat conductivity.
In the first stage, the heat conduction tends to decrease the local Atwood number
and broaden the interfaces of the density and temperature profiles.
In the second stage, part of the compressive energy stored in the fluid is released and
transformed into kinetic energy, thereby accelerating the Rayleigh-Taylor instability.
The local thermodynamic non-equilibrium indicators provide useful observables
to physically track the interfaces.
Indeed, the global or mean thermodynamic non-equilibrium indicators
permit to discriminate the two stages of the Rayleigh-Taylor instability.
In the first stage, the system slowly evolves towards its equilibrium, while in the
later stage, as the interface develops, the system moves away from
local equilibrium, especially in the regions near complex interfaces.
The above behaviour is enhanced by increasing compressibility, and so
are the amplitudes of thermodynamic non-equilibrium kinetic modes. Besides a deeper physical insight into the kinetic procedures, the methodology and resulting observations
may help to formulate more accurate meso and macroscale models for the complex compressible phenomena.

\section*{Acknowledgments}

The authors thank Prof. Hua Li, Drs. Chuandong Lin, Qing Li, Fangbao Tian, Zhipeng Liu and Yudong Zhang for many helpful discussions.
 AX and GZ acknowledge support of Foundation of LCP and National Natural Science Foundation of China (under Grant No. 11475028). HL acknowledges support of National Natural Science Foundation of China (under Grant No. 11301082), China Postdoctoral Science Foundation (under Grant No. 2014M550660) and Natural Science Foundation of Fujian Provinces (under Grant Nos. 2014J05003, JA13069, JB13020). YG acknowledges support of National Natural Science Foundation of China (under Grant No. 11202003) and Natural Science Foundation of Hebei Provinces (under Grant Nos. A2013409003, A201500111).

\appendix
\section{}
\setcounter{figure}{0}%
\renewcommand{\thefigure}{A\arabic{figure}}% The role of these two commands is just an order to reset the count of the following picture.

\begin{figure}[!ht]
\center {
\epsfig{file=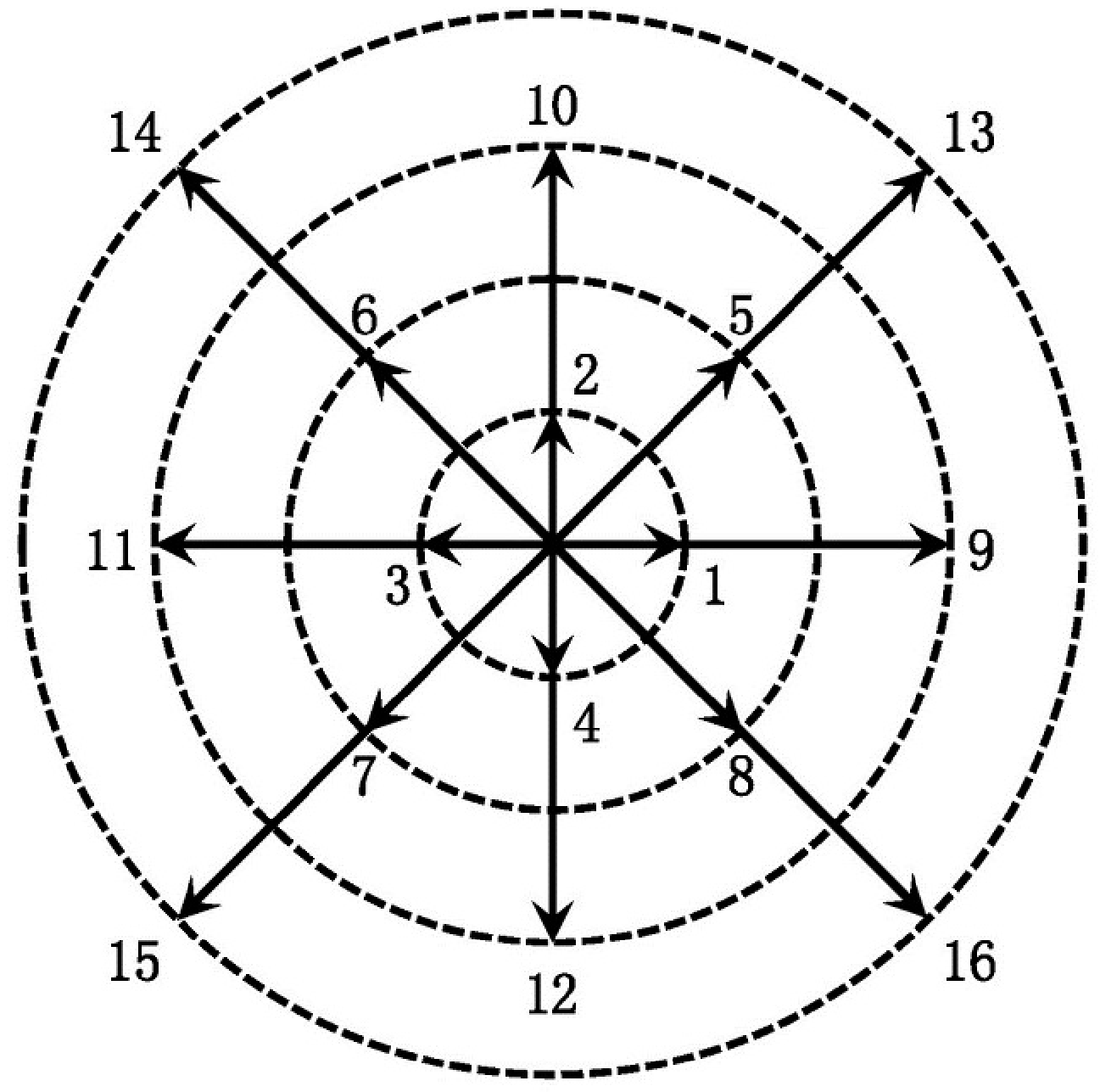,bbllx=120pt,bblly=50pt,bburx=490pt,bbury=420pt,
width=0.3\textwidth,clip=}}
\caption{Sketch of the DVM used in the present paper. The numbers in the figure are the indexes of the discrete velocities.}
\label{FIGA1}
\end{figure}

In this work we
use a 2D 16-velocity (D2V16) model schematically shown in Fig. \ref{FIGA1}. ($|\mathbf{v}_{i}|=c$ for $i=1,\cdots,4$, $|\mathbf{v}_{i}|=2c$ for $i=5,\ldots,8$, $|\mathbf{v}_{i}|=3c$ for $9,\cdots,12$, $|\mathbf{v}_{i}|=4c$ for $i=13,\cdots,16$. $\eta_i=\eta_0$ for $i=1,\cdots,4$, and $\eta_i=0$ for $i=5,\cdots,16$.)

%%%%%%%%%%%%%%%%%%%%%%%%%%%%%%%%%%%%%%%%%%%%%%%%%%%%%%%%%%%%%%%%%%%%%%%%%%%%%%%%
%%%%%%%%%%%%%%%%%%%%%%%%%%%%%%%%%%%%%%%%%%%%%%%%%%%%%%%%%%%%%%%%%%%%%%%%%%%%%%%%%

The equilibrium distribution function $f_{i}^{eq}$ satisfy the following seven kinetic moments \cite{Watari2004}
\begin{eqnarray}\label{mo1}
\mathbf{M}_{0}(f_{i}^{eq})=\sum\limits_{i}f_{i}^{eq}=\rho,
\end{eqnarray}
\begin{eqnarray}\label{mo2}
\mathbf{M}_{1}(f_{i}^{eq})=\sum\limits_{i}f_{i}^{eq}\mathbf{v}_i=\rho \mathbf{u},
\end{eqnarray}
\begin{eqnarray}\label{mo3}
\mathbf{M}_{2,0}(f_{i}^{eq})=\sum\limits_{i}f_{i}^{eq}(\mathbf{v}_{i}\cdot \mathbf{v}
_{i}+\eta_{i}^2)=\rho\big[(D+n)T+\mathbf{u}\cdot\mathbf{u}\big],
\end{eqnarray}
\begin{eqnarray}\label{mo4}
\mathbf{M}_{2}(f_{i}^{eq})=\sum\limits_{i}f_{i}^{eq}\mathbf{v}_{i} \mathbf{v}_{i}=\rho (T\mathbf{I} +\mathbf{u}\mathbf{u}),
\end{eqnarray}
\begin{eqnarray}\label{mo5}
\mathbf{M}_{3,1}(f_{i}^{eq})=\sum\limits_{i}f_{i}^{eq}(\mathbf{v}_{i}\cdot \mathbf{v}
_{i}+\eta_{i}^2)\mathbf{v}_{i}
=\rho \mathbf{u}\big[(D+n+2)T+\mathbf{u}\cdot\mathbf{u}\big],
\end{eqnarray}
\begin{eqnarray} \label{mo6}
\mathbf{M}_{3}(f_{i}^{eq})=\sum\limits_{i}f_{i}^{eq}\mathbf{v}_{i}\mathbf{v}_{i}
\mathbf{v}_{i}=\rho\big[T(\mathbf{u}_\alpha \mathbf{e}_\beta \mathbf{e}_\chi\delta_{\beta\chi}
+\mathbf{u}_\beta\mathbf{e}_\alpha\mathbf{e}_\chi\delta_{\alpha\chi}
+\mathbf{u}_\chi\mathbf{e}_\alpha\mathbf{e}_\beta\delta_{\alpha\beta})
+\mathbf{u}\mathbf{u}\mathbf{u}\big],
\end{eqnarray}
\begin{eqnarray}\label{mo7}
\begin{array}{c}
\mathbf{M}_{4,2}(f_{i}^{eq})=\sum\limits_{i}f_{i}^{eq}(\mathbf{v}_{i}\cdot \mathbf{v}
_{i}+\eta_{i}^2)\mathbf{v}_{i}\mathbf{v}_{i} \\[10pt]
=\rho T \big[(D+n+2)T
+\mathbf{u}\cdot\mathbf{u}\big]\mathbf{e}_\alpha\mathbf{e}_\beta\delta_{\alpha\beta} % \notag %\nonumber\\
+\rho \mathbf{u}\mathbf{u}\big[(D+n+4)T+\mathbf{u}\cdot\mathbf{u}\big],
\end{array}
\end{eqnarray}
where $D$ is the space dimension. Besides the translational degrees of freedom, $\eta$ is a free parameter introduced to describe the $n$ extra degrees of freedom corresponding to the molecular rotation and/or vibration.
$\mathbf{M}^*_{m,n}(\cdot)$ stands for that the $m$-th order tensor is contracted to a $n$-th order one. For 2D case, $\mathbf{M}^*_{0}$ and $\mathbf{M}^*_{2,0}$ are scalars; $\mathbf{M}^*_{1}$ and $\mathbf{M}^*_{3,1}$ are vectors; $\mathbf{M}^*_{2}$ and $\mathbf{M}^*_{4,2}$ are the 2nd order tensors; $\mathbf{M}^*_{3}$ is a 3rd order tensor. The trace of moment $\mathbf{M}^*_{2}$ is a conserved quantity and describes the energy, and its off-diagonal components correspond to the shear effects, which may not be zero whenever the system is in non-thermodynamic equilibrium state \cite{XuLin2014}.
From the Chapman-Enskog analysis, such a DBM can recover the following Navier-Stokes equations
\begin{eqnarray}\label{A8}
\left\{
\begin{array}{ll}
\partial_t \rho+\nabla\cdot (\rho \mathbf{u})=0, &  \\[10pt]
\partial_t(\rho \mathbf{u})+\nabla\cdot(P\mathbf{I}+\rho \mathbf{u} \mathbf{u})-\nabla \cdot \mathbf{P}^{\prime}=\rho \mathbf{a}, &  \\[10pt]
\partial_t\Big[\rho\Big(e+\dfrac{1}{2}|\mathbf{u}|^2\Big)\Big]
+\nabla\cdot\Big[\rho\Big(e+\dfrac{1}{2}|\mathbf{u}|^2\Big)\mathbf{u}+P\mathbf{u}\Big]\\[10pt]
=\nabla\cdot\Big(\kappa\nabla T+\mathbf{u}\cdot \mathbf{P}^{\prime}\Big)+\rho \mathbf{u}\cdot\mathbf{a}, &
\end{array}
\right.
\end{eqnarray}
with
\begin{eqnarray}\label{A9}
\mathbf{P}^{^{\prime }}=\mu\Big[\nabla \mathbf{u}+(\nabla \mathbf{u})^T-\dfrac{2}{D+n} \nabla\cdot\mathbf{u}\mathbf{I}\Big],
\end{eqnarray}
in the continuum limit,
where
\begin{eqnarray}\label{A10}
P=\rho T,~~e=\dfrac{D+n}{2}T
\end{eqnarray}
stand for the pressure and the total internal energy, respectively.
$\mu$, $\kappa$ are the dynamic viscosity and thermal conductivity coefficients having the following relations with $\tau$
\begin{eqnarray}\label{A11}
\mu = P \tau,~~\kappa=c_p P \tau
\end{eqnarray}
where $c_p$ is the specific-heat at constant pressure.

\end{document}